\documentclass[a4paper,12pt]{article}
\pdfoutput=1
\usepackage{jheppub} 
\usepackage{amsmath,amssymb}
\usepackage{graphicx}
\usepackage{float}
\usepackage[export]{adjustbox}
\usepackage[utf8]{inputenc}
\usepackage{slashed}
\usepackage{bbold}
\usepackage{xr}
\usepackage{tikz}
\usetikzlibrary{arrows,shapes}
\usepackage{subfig}
\usepackage[utf8]{inputenc}
\usepackage{simpler-wick}
\usepackage{comment}
\usepackage{amssymb}
\usepackage{amsmath}
\usepackage{amsfonts}
\usepackage{array}
\usepackage{multirow}
\usepackage{psfrag}
\usepackage{xcolor}
\graphicspath{{figures/}}
\usepackage{leftindex}

\def\bea{\begin{eqnarray}}
\def\eea{\end{eqnarray}}
\def\tens{\ensuremath{T}}
\newcommand{\vev}[1]{\ensuremath{ \left< {#1} \right>}}
\newcommand{\TO}[1]{\ensuremath{{\bf {\cal T}} \left\{ {#1} \right\}}}

\newcommand{\braket}[2]{\ensuremath{u^{ {#1}  {#2}}}}

\newcommand{\comm}[2]{\left[ {#1}, {#2} \right] }

\makeatletter
\def\overbracket#1{\mathop{\vbox{\ialign{##\crcr\noalign{\kern3\p@}
\downbracketfill\crcr\noalign{\kern3\p@\nointerlineskip}
$\hfil\displaystyle{#1}\hfil$\crcr}}}\limits}
\def\underbracket#1{\mathop{\vtop{\ialign{##\crcr
$\hfil\displaystyle{#1}\hfil$\crcr\noalign{\kern3\p@\nointerlineskip}
\upbracketfill\crcr\noalign{\kern3\p@}}}}\limits}
\def\overparenthesis#1{\mathop{\vbox{\ialign{##\crcr\noalign{\kern3\p@}
\downparenthfill\crcr\noalign{\kern3\p@\nointerlineskip}
$\hfil\displaystyle{#1}\hfil$\crcr}}}\limits}
\def\underparenthesis#1{\mathop{\vtop{\ialign{##\crcr
$\hfil\displaystyle{#1}\hfil$\crcr\noalign{\kern3\p@\nointerlineskip}
\upparenthfill\crcr\noalign{\kern3\p@}}}}\limits}
\def\downparenthfill{$\m@th\braceld\leaders\vrule\hfill\bracerd$}
\def\upparenthfill{$\m@th\bracelu\leaders\vrule\hfill\braceru$}
\def\upbracketfill{$\m@th\makesm@sh{\llap{\vrule\@height3\p@\@width.7\p@}}%
\leaders\vrule\@height.7\p@\hfill
\makesm@sh{\rlap{\vrule\@height3\p@\@width.7\p@}}$}
\def\downbracketfill{$\m@th
\makesm@sh{\llap{\vrule\@height.7\p@\@depth2.3\p@\@width.7\p@}}%
\leaders\vrule\@height.7\p@\hfill
\makesm@sh{\rlap{\vrule\@height.7\p@\@depth2.3\p@\@width.7\p@}}$}
\makeatother

\title{Tensor reduction of loop integrals}

\author[]{Charalampos Anastasiou,}
\author[]{Julia Karlen,}
\author[]{Matilde Vicini}
\affiliation{Institute for Theoretical Physics, ETH Zurich, 8093
  Z\"urich, Switzerland}
\emailAdd{babis@phys.ethz.ch}
\emailAdd{karlenj@phys.ethz.ch}
\emailAdd{mvicini@phys.ethz.ch}

\abstract{ 
The computational cost associated with reducing tensor integrals to scalar integrals using the Passarino-Veltman method is dominated by the diagonalisation of large systems of equations. These systems of equations are sized according to the number of independent tensor elements that can be constructed using the metric and external momenta.

In this article, we present a closed-form solution of this diagonalisation problem in arbitrary tensor integrals. We employ a basis of tensors whose building blocks are the external momentum vectors and a metric tensor transverse to the space of external momenta. The scalar integral coefficients of the basis tensors are obtained by mapping the basis elements to the elements of an orthogonal
{\it dual basis}. This mapping is succinctly expressed through a formula that resembles the ordering of operators in Wick's theorem.

Finally, we provide examples demonstrating the application of our tensor reduction formula to Feynman diagrams in QCD $2 \to 2$ scattering processes, specifically up to three loops.
 }


\begin{document}
\preprint{}
\maketitle 

\section{Introduction}
\label{sec:Introduction}

In gauge and gravity theories, Feynman diagrams give rise to tensor integrals. 
Practical computations are often organised by first projecting these tensor integrals onto scalar integrals. For the latter, powerful integration by parts methods \cite{TKACHOV198165,CHETYRKIN1981159,Laporta:2001dd}
can be employed, reducing them to linear combinations of a smaller set of master integrals.

Passarino and Veltman introduced a general technique for the reduction of tensor integrals to scalar integrals in their seminal publication of Ref.~\cite{Passarino:1978jh}. Their method has found countless applications since its invention. However, it involves inverting systems of equations which rapidly become intractable as the rank of the tensors increases. This poses a challenge for computations of scattering amplitudes for processes with a large number of external particles and at high loop orders.

Various techniques have been developed as alternatives to the Passarino-Veltman method. 
At one-loop, the symmetric structure of tensor integrals has enabled the development of very efficient reduction methods~\cite{Ezawa:1990dh,Denner:2005nn,Binoth:2008uq,vanHameren:2009vq,vanHameren:2009dr,Cascioli:2011va}. In addition, at one-loop, a refined understanding of the integrand structure of gauge theory amplitudes~\cite{delAguila:2004nf,Britto:2004nc,Britto:2006sj,Ossola:2006us,Forde:2007mi,Ellis:2007br,Giele:2008bc,Lazopoulos:2008ex,Berger:2008sj,Ellis:2008ir,Berger:2009zg,Berger:2009zb,Britto:2010xq,Ellis:2011cr,Actis:2012qn} has revolutionised their reduction to master integrals both at a conceptual and at a practical level. An ambitious research programme at two loops has already lead to remarkable breakthroughs~\cite{Ita:2015tya,Abreu:2017hqn,Badger:2017jhb,Abreu:2020lyk,Abreu:2020xvt,Abreu:2021asb,Badger:2022ncb,Abreu:2023bdp,Badger:2023mgf}. The aforementioned tensor integral and amplitude reduction methods, not only achieve a reduction to scalar integrals, but they also achieve a reduction to master integrals.

Other methods which cast arbitrary tensor integrals as generic scalar integrals (that, in turn, need to be further reduced with integration by parts or other methods to master integrals) exploit the functional form of momentum and parametric representations~\cite{Tarasov:1996br,Anastasiou:1999bn} of tensor integrals. For amplitudes, methods to select external states of definite helicity can be very efficient, 
as it has been shown in the recent works of Ref.~\cite{Chen:2019wyb} and Refs.~\cite{Peraro:2019cjj,Peraro:2020sfm}. For unpolarised scattering, one can avoid a reduction of tensors to scalar integrals by computing squared amplitudes summed over spins and polarisations of external particles, as they appear in cross-sections.

 The Passarino-Veltman method has been superseded by all such other methods in a broad range of applications within particle physics phenomenology. For instance, as far as we are aware, no direct computation of tensor integrals in  amplitudes for scattering processes involving four or more external particles has been pursued with a Passarino-Veltman reduction beyond one loop. Such challenging computations have been carried out with alternative methods (see, for example,  \cite{Garland:2002ak,Bern:2002tk,Caola:2022dfa,Badger:2023xtl,Gehrmann:2023jyv,Gehrmann:2023zpz}).
However, we will show that the Passarino-Veltman method can be simplified. We believe that the improved method is competitive for cutting-edge multi-loop computations in perturbation theory.

An efficient Passarino-Veltman reduction method is valuable for a wide range of applications. This method can be used to disentangle loop integrations and spinor algebra, with the possibility of carrying out integrals fully in $D=4-2\epsilon$ dimensions. This, in turn, enables the treatment of four-dimensional external states and $\gamma_5$ in various prescriptions and regularization schemes flexibly. With the Passarino-Veltman method, one can tackle tensor integrals that do not necessarily originate from contractions with spin gamma matrices. Such tensor integrals emerge, for example, in asymptotic expansions around various kinematic limits, such as small momenta. Tensor integrals also emerge in fields of physics beyond particle physics (see an example in the evaluation of cosmological correlators for the Large Scale Structure of the universe in Ref.~\cite{Anastasiou:2022udy}).

In this article, we present a closed-form solution to the inversion problem in the Passarino-Veltman reduction. Our derivation is based on the observation that all possible tensor structures which may emerge in the tensor reduction of a general tensor integral can be identified through a properly defined ordering of operators associated with tensor indices. 
We project tensor integrals onto scalar integrals with a compact formula that comprises three structures:
\begin{enumerate}
\item Metric tensors which are transverse to external momenta. 
\item Tensors  which are orthogonal to products of transverse metric tensors. These have been introduced in Ref.~\cite{Ruijl:2018poj} for the tensor reduction of multi-loop tadpole integrals. 
\item Linear combinations of momenta that are orthogonal to the external momenta. This construction was originally developed  by van Neerven and Vermaseren in Ref.~\cite{vanNeerven:1983vr} for the reduction of four-dimensional 
one-loop integrals with five or more external particles to box integrals. More recently, it has been invoked in methods for   reduction to master integrals of one-loop amplitudes, see for example Refs.~\cite{Ossola:2006us,Ellis:2011cr}, and 
two-loop amplitudes with the method of Ref.~\cite{Ita:2015tya}.
It has also been used for the projection to helicity amplitudes in Ref.~\cite{Peraro:2019cjj}. 
\end{enumerate}
All terms of the projection are generated efficiently in a closed form, with operations analogous to contractions in applying Wick's theorem for the time-ordering of bosonic free field operators. 

Our method provides an analytical solution to the large set of equations in the Passarino-Veltman system. This tensor reduction technique is universally applicable, being identical for all integrals with the same tensor rank and external momenta. Moreover, it remains independent of other integral-specific characteristics, such as the integral's topology, at fixed rank and number of external momenta.

Our article is organised as follows. In Section~\ref{sec:problem_TR}, we describe the relation between conventionally used tensor elements, which form a basis for the Passarino-Veltman reduction, and an ordering of operators that we attribute to the tensor indices. In Section~\ref{sec:dualbases}, we introduce a metric that is transverse to the space of external momenta and dual elements for external momentum vectors and products of transverse metrics. With these ingredients, we build two dual bases which we use for the tensor reduction. We present our main result for the Passarino-Veltman reduction of generic tensor integrals in closed form in Section~\ref{sec:ingredients}. In Section~\ref{sec:applications}, we demonstrate the application of our formula to tensors up to rank seven, corresponding to QCD Feynman diagrams with four external partons through a perturbative order of three loops. Finally, we summarise our work and present our conclusions in Section~\ref{sec:conclusions}.

\section{Algebraic analogy of tensor reduction and the application of Wick's theorem in ordering operators}
\label{sec:problem_TR}

A general tensor loop integral of rank $R$ in $D$ space-time dimensions
has the form
\begin{eqnarray}
I^{\mu_1 \ldots \mu_R} \equiv \int d^Dk_1\ldots d^Dk_L \,
  \tau^{\mu_1\ldots \mu_R}\left(\{ k_i \} \right) f\left( \{ k_i\},
  p_1,\ldots ,  p_{N_{\text{p}}} \right) \, . 
\end{eqnarray}
$f$ is a scalar function which depends on $N_{\text{p}}$ independent external
momenta $p_i$, and $\tau$ is a rank-$R$ tensor which is a linear combination of products of the loop momenta. It can be easily seen, for example~\cite{Anastasiou:1999bn} in Feynman or Schwinger parameter representations, that tensor integrals can be written as a superposition of all independent rank-$R$ tensors which can be built by multiplying external momentum vectors and/or the metric. Generically, we write 
\begin{equation}
  \label{eq:tensordecomposition}
I^{\mu_1 \ldots \mu_R}  = \sum_a I_a \,\tens_a^{\mu_1 \ldots \mu_R}\left(
  \eta^{\alpha \beta}, p_i^{\alpha} \right).
\end{equation}
The tensors $\tens_a^{\mu_1 \ldots \mu_R}$ represent all such independent 
possible tensors of rank $R$ of the metric and external momenta. 
The coefficients $I_a$ are scalar
integrals which are amenable to further reduction with techniques
such as integration by parts~\cite{TKACHOV198165,CHETYRKIN1981159,Laporta:2001dd}. 

Expressing the tensor integral as a superposition of scalar integrals and deriving the 
coefficients $I_a$ is an often daunting linear algebra task. 
For a rank-$R$ tensor and $N_{\rm p}$ independent external momenta, tensor reduction generates 
\begin{eqnarray}
\label{eq:Ntensors}
N_{\rm tensors} = \sum_{n=0}^{\lfloor \frac R 2 \rfloor} \frac{R!}{2^n \, n! (R-2 n)!} \, N_{\rm p}^{R-2 n}
\end{eqnarray}
terms in total, where $\lfloor x\rfloor = \max\{m\in\mathbb{Z} \vert m \leq x\}$ is the floor function. For example, in two-loop $2 \to 2$ scattering amplitudes we require $R=5, N_{\text{p}}=3$, which results in 558 terms. For $R=6, N_{\text{p}}=4$ in $2 \to 3$ two-loop processes we have 8671 terms and, by adding one more loop, for $R=8, N_{\text{p}}=4$ we have 240809 terms. In regular Passarino-Veltman reduction, this produces a system of $N_{\rm tensors}$ equations that requires the inversion of an $N_{\rm tensors} \times N_{\rm tensors}$ matrix in order to derive the coefficients $I_a$.  

There is an algebraic way to list all the tensor elements (products of metric tensors and external momentum vectors) which appear in the typical basis of Passarino-Veltman reduction. 
For concreteness, let us consider the outcome of tensor reduction of a rank-4 tensor integral. This has the form
\begin{align}
\label{eq:R4tensordecomposition}
I^{\mu_1 \mu_2 \mu_3 \mu_4} =& 
\sum_{i_1,i_2,i_3,i_4=1}^{N_{\rm p}} c^{(0)}_{i_1i_2i_3i_4} \,  
p_{i_1}^{\mu_1} p_{i_2}^{\mu_2}
p_{i_3}^{\mu_3} p_{i_4}^{\mu_4} 
\nonumber \\ & \hspace{-1cm}
+\eta^{\mu_1 \mu_2 } \sum_{i_3,i_4=1}^{N_{\rm p}}   c^{(12)}_{i_3i_4} \,  
p_{i_3}^{\mu_3} p_{i_4}^{\mu_4}
+\eta^{\mu_1 \mu_3 } \sum_{i_2,i_4=1}^{N_{\rm p}}   c^{(13)}_{i_2i_4} \,  
p_{i_2}^{\mu_2} p_{i_4}^{\mu_4}
+\eta^{\mu_1 \mu_4 } \sum_{i_2,i_3=1}^{N_{\rm p}}   c^{(14)}_{i_2i_3} \,  
p_{i_2}^{\mu_2} p_{i_3}^{\mu_3} 
\nonumber \\ & \hspace{-1cm}
+\eta^{\mu_2 \mu_3 } \sum_{i_1,i_4=1}^{N_{\rm p}}   c^{(23)}_{i_1i_4} \,  
p_{i_1}^{\mu_1} p_{i_4}^{\mu_4}
+\eta^{\mu_2 \mu_4 } \sum_{i_1,i_3=1}^{N_{\rm p}}   c^{(24)}_{i_1i_3} \,  
p_{i_1}^{\mu_1} p_{i_3}^{\mu_3}
+\eta^{\mu_3 \mu_4 } \sum_{i_1,i_2=1}^{N_{\rm p}}   c^{(34)}_{i_1i_2} \,  
p_{i_1}^{\mu_1} p_{i_2}^{\mu_2}
\nonumber \\ &  \hspace{-1cm}
+c^{(12,34)} \eta^{\mu_1 \mu_2 }
\eta^{\mu_3 \mu_4 }
+ c^{(13,24)} \eta^{\mu_1 \mu_3 }
\eta^{\mu_2 \mu_4 }
+ c^{(14,23)} \eta^{\mu_1 \mu_4 }
\eta^{\mu_2 \mu_3 } \, . 
\end{align}
We would like to generate the list of all tensors 
\begin{eqnarray}
\label{eq:R4elements}
\left\{ 
p_{i_1}^{\mu_1} p_{i_2}^{\mu_2}p_{i_3}^{\mu_3} p_{i_4}^{\mu_4}, \, 
p_{i_1}^{\mu_1} p_{i_2}^{\mu_2} \eta^{\mu_3 \mu_4 }, \, 
\ldots , 
\eta^{\mu_1 \mu_2 } \eta^{\mu_3 \mu_4 } ,  \, \ldots 
\right\}
\end{eqnarray}
 in the right-hand side of Eq.~\eqref{eq:R4tensordecomposition}. 
 
We consider all independent vectors $p_i^{\mu_a}$ and we associate to their sum an \textit{index operator} ${\bf a}$ which indicates the index  $\mu_a$, 
\begin{eqnarray}
    \sum_{i=1}^{N_{\rm p}} p_i^{\mu_a} \to {\bf a} \, . 
\end{eqnarray}
This assignment defines an ordering of the indices $\mu_a$. Explicitly, 
\begin{eqnarray}
    \sum_{i=1}^{N_{\rm p}} p_i^{\mu_1} \to {\bf 1}\, , \; 
    \sum_{i=1}^{N_{\rm p}} p_i^{\mu_2} \to {\bf 2} \, ,  \; 
    {\rm etc.}
\end{eqnarray}
We now attribute to the index operators a \textit{commutator}
\begin{eqnarray}
\label{eq:commutator}
    \comm{\bf a}{\bf b} \equiv 
    \wick{\c {\bf a} \c {\bf b}}
    = \eta^{\mu_a \mu_b}
\end{eqnarray}
and an \textit{ordering operation} 
\begin{eqnarray}
{\cal T} \left( {\bf a_1} {\bf a_2} \cdots \right),
\end{eqnarray}
where, using the commutation of Eq.~\eqref{eq:commutator}, we bring operators ${\bf a_i}$ with a larger tensor-index label to the left, in front of operators with a smaller tensor-index label.  For example, 
\begin{eqnarray}
\label{eq:T4primitive}
{\cal T} 
\left({\bf 1} {\bf 2} {\bf 3} {\bf 4}\right)   &=& 
{\cal T} \left({\bf 2} {\bf 1} {\bf 3} {\bf 4}\right)
+ \wick{\c {\bf 1} \c {\bf 2}} \,  
{\cal T} \left({\bf 3} {\bf 4}\right)
\nonumber \\ 
&=& {\cal T} \left({\bf 2} {\bf 3} {\bf 1} {\bf 4}\right)
+ \wick{\c {\bf 1} \c {\bf 3}} \,  
{\cal T} \left({\bf 2} {\bf 4}\right)
+ \wick{\c {\bf 1} \c {\bf 2}} \,  
\left({\bf 4} {\bf 3}\right)
+ \wick{\c {\bf 1} \c {\bf 2}} \,
 \wick{\c {\bf 3} \c {\bf 4}} 
 \nonumber \\ 
 &=& 
 {\cal T} \left({\bf 2} {\bf 3} {\bf 4} {\bf 1}\right)
+\wick{\c {\bf 1} \c {\bf 4}} \,  
{\cal T} \left({\bf 2} {\bf 3}\right)
 + \wick{\c {\bf 1} \c {\bf 3}} \,  
\left({\bf 4} {\bf 2}\right)
+ \wick{\c {\bf 1} \c {\bf 3}}  
 \wick{\c {\bf 2} \c {\bf 4}}
 \nonumber \\ && 
+ \wick{\c {\bf 1} \c {\bf 2}} \,  
 \left({\bf 4} {\bf 3}\right)
+ \wick{\c {\bf 1} \c {\bf 2}} \,
 \wick{\c {\bf 3} \c {\bf 4}}
 \nonumber \\ 
 &=& \ldots 
 \nonumber \\ 
 &=& \left({\bf 4} {\bf 3} {\bf 2} {\bf 1}\right)
+\wick{\c {\bf 1} \c {\bf 2}} \, \left({\bf 4} {\bf 3}\right)
+\wick{\c {\bf 1} \c {\bf 3}} \, \left({\bf 4} {\bf 2}\right)
+\wick{\c {\bf 2} \c {\bf 3}} \, \left({\bf 4} {\bf 1}\right) 
\nonumber \\ &&
+\wick{\c {\bf 1} \c {\bf 4}} \, \left({\bf 3} {\bf 2}\right)
+\wick{\c {\bf 2} \c {\bf 4}} \, \left({\bf 3} {\bf 1}\right)
+\wick{\c {\bf 3} \c {\bf 4}} \, \left({\bf 2} {\bf 1}\right)
\nonumber \\ &&
+ \wick{\c {\bf 1} \c {\bf 2}} \,
 \wick{\c {\bf 3} \c {\bf 4}}
 + \wick{\c {\bf 1} \c {\bf 3}} \,
 \wick{\c {\bf 2} \c {\bf 4}}
 + \wick{\c {\bf 1} \c {\bf 4}} \, 
 \wick{\c {\bf 2} \c {\bf 3}} \, . 
\end{eqnarray}
The last equation is the sum of all rank-4 tensors in the list of Eq.~\eqref{eq:R4elements}, as it can be seen by substituting the contractions as metric tensors and index operators with the sum of momentum vectors. 
Generally, we can use the ordering, 
\begin{eqnarray}
{\cal T}\left({\bf 1} {\bf 2} \cdots  {\bf R} \right) \, ,    
\end{eqnarray}
to sum up all the elements of the basis in which we can express a tensor integral  $I^{\mu_1 \mu_2 \ldots \mu_R}$ of rank $R$.

In the following sections, inspired by the combinatorial resemblance of tensor reduction and the ordering of operators with Wick contractions, we will go one step further. We will develop a new ordering operation which will yield the full answer for tensor reduction. That is, we will obtain the sum of all the basis tensor-elements weighted with the correct scalar integral coefficients (such as the $c^{(X)}_{\ldots}$ coefficients which appear in the right-hand side of  Eq.~\eqref{eq:R4tensordecomposition}).

\section{Tensor reduction and dual bases}
\label{sec:dualbases}

Our starting point is to construct two complete  bases of tensors, $\mathcal{B} = \{ \tens_{a}^{\mu_1 \ldots \mu_R} \}$ and $\vev{\mathcal{B}} =\{\vev{\tens_{a}}^{\mu_1 \ldots \mu_R} \}$, which are \textit{dual}, satisfying the property 
\begin{eqnarray}
  \label{eq:Tbasis}
\vev{\tens_{a}} \cdot \tens_b \equiv  \vev{\tens_{a}}^{\mu_1 \ldots \mu_R}  \tens_{b,
  \mu_1 \ldots \mu_R}  = \delta_{ab} \,.  
\end{eqnarray}
The two bases do not need to be, and we will not take them to be, individually orthonormal.

In what follows, we will present an explicit construction of the dual bases. 
Assuming their existence, the reduction of tensor integrals takes a simple form.  
Indeed, by contracting both sides of Eq.~\eqref{eq:tensordecomposition} with
$\vev {\tens_{b}}$ we can determine the scalar coefficients $I_b$.
We obtain
\begin{equation}
  \label{eq:TRa}
I^{\mu_1 \ldots \mu_R}  =   I_ {\alpha_1 \ldots \alpha_R} \, \sum_{a} \,
\vev{ \tens_a}^{\alpha_1 \ldots \alpha_R} \,  \tens_a^{\mu_1 \ldots \mu_R}\,.
\end{equation}
Alternatively, we can also write
\begin{equation}
  \label{eq:TRb}
I^{\mu_1 \ldots \mu_R}  =   I_ {\alpha_1 \ldots \alpha_R} \, \sum_{a}\, 
 \tens_a^{\alpha_1 \ldots \alpha_R} \, 
  \vev{\tens_a}^{\mu_1 \ldots \mu_R} \, . 
\end{equation}
Notice that on the right-hand side of Eqs.~\eqref{eq:TRa}-\eqref{eq:TRb} the
tensor indices of the integral are contracted and the tensor integral is reduced to scalar integrals. 

 We will now build the dual bases which are needed for materialising Eq.~\eqref{eq:TRa} and Eq.~\eqref{eq:TRb}. 

\subsection{Dual tensors for products of external momenta}

Let us first start with a basis of $N_{\text{p}}$ external momentum vectors $p_i^\mu, \; i=1\ldots N_{\text{p}}$. 
To find their dual, we adopt the method of van Neerven and Vermaseren in Ref.~\cite{vanNeerven:1983vr} (see, also, Section 3 of Ref.~\cite{Ellis:2011cr} for a elucidating variation of the formalism). We first form the symmetric $N_{\text{p}} \times N_{\text{p}}$ matrix of scalar products, 
\begin{equation}
  \label{eq:Pi-matrix}
\Pi_{ij} \equiv p_i \cdot p_j \,, 
\end{equation}
and, with linear algebra methods, we find the inverse matrix
$\Delta_{ij}$, 
\begin{equation}
\sum_{k=1}^{N_{\text{p}}} \Pi_{ik} \Delta_{kj} = \delta_{ij}. 
\end{equation}
The linear combinations
\begin{equation}
  \label{eq:q-momenta}
  \vev{p_i}^\mu \equiv \sum_{j=1}^{N_{\text{p}}} \Delta_{ij} \, p_j^\mu
\end{equation}
are dual to $p_i$ in the sense of Eq.~\eqref{eq:Tbasis}. Indeed,  
\begin{eqnarray}
\label{eq:mom_dual_prop}
p_i \cdot \vev{p_j}  = \sum_{k=1}^{N_{\text{p}}} \Delta_{kj} \, p_{i} \cdot p_k  =
  \sum_{k=1}^{N_{\text{p}}} \Pi_{ik}
  \Delta_{kj} = \delta_{ij} \, . 
\end{eqnarray}
We note that
\begin{eqnarray}
\Delta_{ij} = \vev{p_i} \cdot \vev{p_j} \, . 
\end{eqnarray}

We can now easily extend the construction to products of momenta. 
For a rank-$R$ tensor product of momentum vectors
\begin{equation}
T_{\rm p}^{\mu_1 \ldots \mu_R} ={p_{i_1}}^{\mu_1} \ldots p_{i_R}^{\mu_R},  
\end{equation}
we construct its dual tensor as the product of the duals of momentum vectors 
\begin{equation}
\vev{T_{\rm p}}^{\mu_1 \ldots \mu_R} =\vev{{p_{i_1}}^{\mu_1} \ldots p_{i_R}^{\mu_R}  } =\vev{p_{i_1}}^{\mu_1} \ldots \vev{p_{i_R}}^{\mu_R} , 
\end{equation}
which satisfies 
\begin{equation}
T_{{\rm p}, \mu_1 \ldots \mu_R}  \vev{T_{\rm p}}^{\mu_1 \ldots \mu_R} =1 \, . 
\end{equation}
The tensor $T_{\rm p}^{\mu_1 \ldots \mu_R}$ is annihilated by any other
product of $R$ vectors $\vev{p_i}^{\mu}$ than $\vev{T_{\rm p}}^{\mu_1 \ldots \mu_R}$.  However, it is not annihilated by contractions with the metric tensor, such that the condition in Eq.~\eqref{eq:Tbasis} is not fulfilled. To resolve this issue we construct metric tensors as well as their duals, which are transverse to the external momenta. It will be useful to form a rank-2 tensor from the external momenta $p_i$ and their orthogonal momenta $\vev{p_i}$,
\begin{equation}
  \label{eq:bracket}
\braket{\mu}{\nu } \equiv \sum_{i=1}^{N_{\text{p}}} p_i^\mu  \vev{p_i}^\nu = \sum_{i,j=1}^{N_{\text{p}}} \Delta_{ij} p_i^\mu p_j^\nu \, , 
\end{equation}
which acts as the unity  in the space of external momenta. Indeed, we have that 
\begin{eqnarray}
    p_{k, \mu} \braket{\mu}{\nu} =
    \sum_{i,j=1}^{N_{\text{p}}} \Delta_{ij} \,  p_i \cdot p_k \,  p_j^\nu = \sum_{j}^{N_{\text{p}}} \delta_{jk}
    p_j^\nu =p_k^\nu \, . 
\end{eqnarray}
For this reason, we will refer to $\braket{\mu}{\nu}$ as the {\it unit tensor}. 
The natural choice for a metric transverse to external momenta is the tensor obtained by subtracting $u$ from the metric,
\begin{equation}
\eta_\perp^{\mu \nu} \equiv \eta^{\mu \nu} - \braket{\mu}{\nu} \, . 
\end{equation}  
Explicitly, we have that
\begin{equation}
\eta_\perp^{\mu \nu} p_{i,\mu}  = \eta_\perp^{\mu \nu} \vev{p}_{i,\mu} = 0. 
\end{equation}
In addition, the \textit{transverse metric} satisfies
\begin{eqnarray}
\eta_\perp^{\mu \nu} {\eta_{\perp, \nu}}^\rho = \eta_\perp^{\mu \rho} \,
\end{eqnarray}
and the dimensionality of the transverse metric is
\begin{eqnarray}
D_\perp \equiv \eta_\perp^{\mu \nu} \eta_{\perp, {\mu \nu}} = D-N_{\text{p}}, 
\end{eqnarray}
where $N_{\text{p}}$ is the number of independent external momenta.

\subsection{Dual tensors for products of transverse metrics}
\label{sec:metricproducts}

While the transverse metric tensor and products of it are orthogonal to all external momenta and their duals, they are not orthogonal to other metric products.   
We will now construct dual elements $\vev{T_{a}}$  for  tensors $T_{a}$ which are
products of the transverse metric tensor,
\begin{eqnarray}
  \label{eq:Tmetrics}
  T_{\rm metric}^{\mu_1 \ldots \mu_R}
  = \eta_\perp^{\mu_1 \mu_2} \ldots \eta_\perp^{\mu_{R-1} \mu_R}. 
\end{eqnarray}
Our procedure is analogous to the method in Ref.~\cite{Ruijl:2018poj} for tensor loop integrals with no external momenta. 

At even rank $R$, the set of independent tensors of the form of Eq.~\eqref{eq:Tmetrics} contains 
\begin{align}
    N_{\text{metric}} = \frac{R!}{2^{R/2}(R/2)!}
\end{align}
elements. For rank two, there is only one tensor we can form, $\eta_\perp^{\mu_1 \mu_2}$.
We trivially write its dual element in the orthogonal basis,  
\begin{equation}
\label{eq:dualmetricR2}
\vev{\eta_\perp^{\mu_1 \mu_2}}  = \frac{\eta_\perp^{\mu_1 \mu_2}}{D_\perp} \, ,    
\end{equation}
satisfying
\begin{eqnarray}
\eta_{\perp, \mu_1 \mu_2} \, \vev{\eta_\perp^{\mu_1 \mu_2}} =1 \,. 
\end{eqnarray}

At rank four, we find three independent tensors
\begin{equation}
  \left\{
    \eta_\perp^{\mu_1 \mu_2} \eta_\perp^{\mu_3 \mu_4} \,,
        \eta_\perp^{\mu_1 \mu_3} \eta_\perp^{\mu_2 \mu_4} \,,
        \eta_\perp^{\mu_1 \mu_4} \eta_\perp^{\mu_2 \mu_3} \,
  \right\} \, .   
\end{equation}
To construct the dual tensors in the orthogonal basis, we write an ansatz
\begin{eqnarray}
  \vev{    \eta_\perp^{\mu_1 \mu_2} \eta_\perp^{\mu_3 \mu_4} }
  = a    \,  \eta_\perp^{\mu_1 \mu_2} \eta_\perp^{\mu_3 \mu_4}
  + b \left(
  \eta_\perp^{\mu_1 \mu_3} \eta_\perp^{\mu_2 \mu_4}
  +\eta_\perp^{\mu_1 \mu_4} \eta_\perp^{\mu_2 \mu_3}
  \right) \, , 
\end{eqnarray}
in which we have used the $\mu_1 \leftrightarrow \mu_2$ and $\mu_3
\leftrightarrow \mu_4$ and $\left( \mu_1 , \mu_2\right)
\leftrightarrow \left( \mu_3 , \mu_4\right) $  symmetry.  We determine
the $a,b$ coefficients from
the requirements
\begin{eqnarray}
\vev{    \eta_\perp^{\mu_1 \mu_2} \eta_\perp^{\mu_3 \mu_4} }  \, {    \eta_{\perp, \mu_1
  \mu_2} \eta_{\perp, \mu_3 \mu_4} } = 1\,
  \quad 
  \vev{    \eta_\perp^{\mu_1 \mu_2} \eta_\perp^{\mu_3 \mu_4} }  \, {    \eta_{\perp, \mu_1
  \mu_3} \eta_{\perp, \mu_2 \mu_4} } = 0\, .  
\end{eqnarray}
We obtain
\begin{eqnarray}
\label{eq:dualmetricR4}
 \vev{    \eta_\perp^{\mu_1 \mu_2} \eta_\perp^{\mu_3 \mu_4} }
  &=& \frac{D_\perp+1}{D_\perp (D_\perp-1) (D_\perp+2)}    \,  \eta_\perp^{\mu_1 \mu_2} \eta_\perp^{\mu_3 \mu_4}
      \nonumber \\
  &&
  -\frac{1}{D_\perp (D_\perp-1) (D_\perp+2)} \left(
  \eta_\perp^{\mu_1 \mu_3} \eta_\perp^{\mu_2 \mu_4}
  +\eta_\perp^{\mu_1 \mu_4} \eta_\perp^{\mu_2 \mu_3}
  \right) \, . \nonumber \\
\end{eqnarray}  
Similarly, we can construct the elements of the dual basis for
lengthier products. 
The construction for the dual basis of rank six and eight is described in Appendix \ref{sec:Appendix}. The dual basis for the metric tensors up to rank fourteen are available in an ancillary file, where the coefficients are calculated using \textsc{Form}  \cite{vermaseren2000new,Tentyukov_2010,Kuipers_2013}.
These suffice, for example, for the tensors which emerge in QCD Feynman diagrams with four external partons at six loops. 

One can show a useful identity, that contracting the dual transverse metric of rank $R=2n$ with a transverse metric $\eta_{\perp}^{\mu_i\mu_{i+1}}$ with $i$ odd, the dual metric reduces to a dual transverse metric of rank $R-2$. For example
\begin{equation}
    \eta_{\perp}^{\mu_1\mu_2}\vev{\eta_{\perp,\mu_1\mu_2}\eta_{\perp,\rho_1\rho_2}\ldots\eta_{\perp,\rho_{2n-1}\rho_{2n}}}=\vev{\eta_{\perp,\rho_1\rho_2}\ldots\eta_{\perp,\rho_{2n-1}\rho_{2n}}}.
    \label{eq:metric_contraction_short}
\end{equation}
It follows that contracting the dual transverse metric product with an arbitrary number $m$ transverse metrics (contained in the dual) we reduce the dual transverse metric of rank $2(n+m)$ to a dual of rank $2n$:
\begin{align}
    &\eta_{\perp}^{\mu_1\mu_2}\ldots \eta_{\perp}^{\mu_{2m-1}\mu_{2m}}\vev{\eta_{\perp,\mu_1\mu_2}\ldots \eta_{\perp,\mu_{2m-1}\mu_{2m}}\eta_{\perp,\rho_1\rho_2}\ldots\eta_{\perp,\rho_{2n-1}\rho_{2n}}} \nonumber\\
    & =\vev{\eta_{\perp,\rho_1\rho_2}\ldots\eta_{\perp,\rho_{2n-1}\rho_{2n}}}.
    \label{eq:metric_contraction_long}
\end{align}
The proof of Eq.\eqref{eq:metric_contraction_long} can be found in Appendix \ref{sec:proofmetric}.
Note that the dual of a product of transverse metric tensors does not factorise
\begin{align}
    \vev{
\prod_i \eta_{\perp}^{\mu_i \nu_i}
  } \neq \prod_i \vev{\eta_{\perp}^{\mu_i \nu_i}}.
\end{align}

\section{Tensor reduction of a generic loop integral}
\label{sec:ingredients}

We now have all ingredients to express the reduction of a generic tensor integral in a closed form. 
For a generic tensor integral of rank $R$ we have constructed two bases of tensors which are 
\begin{itemize}
\item a basis $\mathcal{B}=\{\tens_a^{\mu_1\ldots\mu_R}\}$ consisting of all possible products of momenta $p_i^\mu$ and the transverse metric tensor $\eta_\perp^{\mu \nu}$,
\item a dual basis $\vev{\mathcal{B}} =\{\vev{\tens_a}^{\mu_1\ldots\mu_R}\}$ consisting of all possible products of dual momenta $\vev{p_i}^\mu$ and duals of products of transverse metric tensors $\vev{\eta_\perp^{\mu_1 \mu_2}\eta_\perp^{\mu_3 \mu_4} \ldots }$.
\end{itemize}
As an example, the basis as well as the dual basis for $N_{\text{p}}$ independent external momenta and tensors of rank one is given by
\begin{align}
\label{eq:basesR1}
    \mathcal{B} &= \{p_1^{\mu},p_2^{\mu},\ldots,p_{N_{\rm p}}^{\mu}\} = \{p_i^{\mu}\}_{i=1...N_{\rm p}},  \\
   \vev{\mathcal{B}} &= \{\vev{p_1}^{\mu},\vev{p_2}^{\mu},\ldots,\vev{p_{N_{\rm p}}}^{\mu}\} = \{\vev{p_i}^{\mu}\}_{i=1...N_{\rm p}}, \nonumber
\end{align}
for rank two the bases are given by
\begin{align}
\label{eq:basesR2}
   \mathcal{B}&= \left\{ p_i^{\mu}p_j^{\nu}\right\}_{i,j=1...N_{\rm p}} \cup \left\{\eta_\perp^{\mu\nu} \right\}, \\
     \vev{\mathcal{B}} &= \left\{ \vev{p_i}^{\mu}\vev{p_j}^{\nu}\right\}_{i,j=1...N_{\rm p}} \cup \left\{\vev{\eta_\perp^{\mu\nu}} \right\}, \nonumber
\end{align}
whereas for rank three we get
\begin{align}
\label{eq:basesR3}
   \mathcal{B}&= \left\{ p_i^{\mu}p_j^{\nu}p_k^{\rho}\right\}_{i,j,k=1...N_{\rm p}} \cup \left\{\eta_\perp^{\mu\nu} p_i^{\rho}, \eta_\perp^{\nu\rho} p_i^{\mu}, \eta_\perp^{\mu\rho} p_i^{\nu}\right\}_{i=1...N_{\rm p}},\\
     \vev{\mathcal{B}} &= \left\{ \vev{p_i}^{\mu}\vev{p_j}^{\nu}\vev{p_k}^{\rho}\right\}_{i,j,k=1...N_{\rm p}} \cup \left\{\vev{\eta_\perp^{\mu\nu}} \vev{p_i}^{\rho}, \vev{\eta_\perp^{\nu\rho}} \vev{p_i}^{\mu}, \vev{\eta_\perp^{\mu\rho}} \vev{p_i}^{nu}\right\}_{i=1...N_{\rm p}}. \nonumber
\end{align}

The two bases $ \mathcal{B}$ and $ \vev{\mathcal{B}}$ for a general rank-$R$ tensor integral satisfy the Eq.~\eqref{eq:Tbasis}. We can then cast a tensor integral in the forms of Eq.~\eqref{eq:TRa} or Eq.~\eqref{eq:TRb} using our dual bases.  
Let us analyse the sum on the right-hand side of Eq.~\eqref{eq:TRb}, with our initial focus on the terms that solely involve external momenta and do not include any transverse metric tensors. We have 
\begin{eqnarray}
 \sum_{a}\, 
 \tens_a^{\alpha_1 \ldots \alpha_R} \, 
  \vev{\tens_a}^{\mu_1 \ldots \mu_R}   &=&
\sum_{i_1,i_2, \ldots , i_R=1}^{N_{\rm p}}
{p_{i_1}}^{\alpha_1} \ldots p_{i_R}^{\alpha_R}
\vev{
{p_{i_1}}^{\mu_1} \ldots p_{i_R}^{\mu_R}  
} + \left( \mbox{$\eta_\perp$ {\small terms}} \right)
\nonumber \\
&& \hspace{-2cm}
= \sum_{i_1,i_2, \ldots , i_R=1}^{N_{\rm p}}
{p_{i_1}}^{\alpha_1} \ldots p_{i_R}^{\alpha_R}
\vev{p_{i_1}}^{\mu_1} \ldots \vev{p_{i_R}}^{\mu_R}  
+ \left( \mbox{$\eta_\perp$ {\small terms}} \right)
\nonumber \\
&& \hspace{-2cm}
= \prod_{k =1}^R 
\left( 
\sum_{i=1}^{N_{\rm p}} 
p_i^{\alpha_k}
\vev{p_i}^{\mu_k}
\right)
+ \left( 
\mbox{$\eta_\perp$ {\small terms}}
\right)
\nonumber \\ 
&& \hspace{-2cm}
=\prod_{k =1}^R 
\braket{\alpha_k}{\mu_k}
+ \left( 
\mbox{$\eta_\perp$ {\small terms}}
\right).
\end{eqnarray}
In the above, we could collect all momentum-only dependent terms in the tensor reduction of Eq.~\eqref{eq:TRb} into a product of unit tensors $\braket{\alpha}{\mu}$. 

We now turn our attention to terms with the transverse metric.  These are of the form,
\begin{eqnarray}
\label{eq:metrictermsinmapping}
&& \sum_{a}\, 
 \tens_a^{\alpha_1 \ldots \alpha_R} \, 
  \vev{\tens_a}^{\mu_1 \ldots \mu_R} 
  \nonumber \\ 
  && 
\owns
\sum_{n=2}^{2\lfloor R/2\rfloor} \sum_{i_{n+1}, \ldots , i_R=1}^{N_{\rm p}}
\eta_\perp^{\alpha_1 \alpha_2}\cdots 
\eta_\perp^{\alpha_{n-1} \alpha_n} 
p_{i_{n+1}}^{\alpha_{n+1}} 
\cdots p_{i_{R}}^{\alpha_{R}}
\vev{ 
\eta_\perp^{\mu_1 \mu_2}\cdots 
\eta_\perp^{\mu_{n-1} \mu_n} 
p_{i_{n+1}}^{\mu_{n+1}} 
\cdots p_{i_{R}}^{\mu_{R}}
}
\nonumber \\ 
&& 
=
\sum_{n=2}^{2\lfloor R/2\rfloor} \sum_{i_{n+1}, \ldots , i_R=1}^{N_{\rm p}}
\eta_\perp^{\alpha_1 \alpha_2}\cdots 
\eta_\perp^{\alpha_{n-1} \alpha_n} 
p_{i_{n+1}}^{\alpha_{n+1}} 
\cdots p_{i_{R}}^{\alpha_{R}}
\vev{ 
\eta_\perp^{\mu_1 \mu_2}\cdots 
\eta_\perp^{\mu_{n-1} \mu_n} }
\vev{p_{i_{n+1}}}^{\mu_{n+1}} 
\cdots \vev{p_{i_{R}}}^{\mu_{R}
}
\nonumber \\ 
&& 
=\sum_{n=2}^{2\lfloor R/2\rfloor}
\eta_\perp^{\alpha_1 \alpha_2}\cdots 
\eta_\perp^{\alpha_{n-1} \alpha_n} 
\vev{ 
\eta_\perp^{\mu_1 \mu_2}\cdots 
\eta_\perp^{\mu_{n-1} \mu_n} }
\prod_{k=n+1}^R \braket{\alpha_k}{\mu_k}.
\end{eqnarray}
 We define a {\it product of contractions} as
\begin{eqnarray}
\label{eq:productcontractions}
  \prod_i
  \wick{ \c {\braket{\alpha_i}{ \mu_i}} \c
  {\braket{\beta_i}{\nu_i}}}
  \equiv \left( \prod_i \eta_{\perp}^{\alpha_i \beta_i} \right) \vev{
\prod_i \eta_{\perp}^{\mu_i \nu_i}
  }.
\end{eqnarray}
With the contraction symbol of Eq.~\eqref{eq:productcontractions}, 
we cast the terms of Eq.~\eqref{eq:metrictermsinmapping}  as 
\begin{eqnarray}
&& 
\sum_{i_{n+1}, \ldots , i_R=1}^{N_{\rm p}}
\eta_\perp^{\alpha_1 \alpha_2}\cdots 
\eta_\perp^{\alpha_{n-1} \alpha_n} 
p_{i_{n+1}}^{\alpha_{n+1}} 
\cdots p_{i_{R}}^{\alpha_{R}}
\vev{ 
\eta_\perp^{\mu_1 \mu_2}\cdots 
\eta_\perp^{\mu_{n-1} \mu_n} 
p_{i_{n+1}}^{\mu_{n+1}} 
\cdots p_{i_{R}}^{\mu_{R}}
} 
\nonumber \\ 
&& \hspace{1.5cm}
= 
  \prod_{i=1}^{\frac n 2}
  \wick{ \c {\braket{\alpha_{2 i-1}}{ \mu_{2 i-1}}} \c
  {\braket{\alpha_{2 i}}{\mu_{2 i}}}}
\prod_{k=n+1}^R \braket{\alpha_k}{\mu_k}\,.
\end{eqnarray}
Finally, we can sum up all terms in Eq.~\eqref{eq:TRb} compactly as an ordering operation of the unit tensors $\braket{\alpha}{ \mu}$. We define an ordering symbol $\ensuremath{{\bf {\cal T}}}$ as 
\begin{eqnarray}
\TO { A_1 \ldots A_n}   \equiv  A_1 \ldots  A_n + \mbox{all contractions,} 
\end{eqnarray}
where $A_i = \braket{\alpha_i}{ \mu_i}$.
For example, for a rank-$4$ tensor we write
\begin{eqnarray}
\label{eq:wickAA}
  \TO{ A_1 A_2 A_3 A_4 } &\equiv&  A_1 A_2 A_3 A_4
  + \wick{\c {A_1} \c {A_2} {A_3} {A_4}}
  + \wick{\c {A_1}  {A_2} \c {A_3} {A_4}}
  + \wick{\c {A_1}  {A_2} {A_3} \c {A_4}}
  \nonumber \\
  &&
  + \wick{ {A_1} \c {A_2} \c {A_3} {A_4}}
  + \wick{ {A_1} \c {A_2}  {A_3} \c  {A_4}}
  + \wick{ {A_1} {A_2} \c {A_3} \c {A_4}}
     \nonumber \\ &&
  + \wick{ \c1 {A_1} \c1  {A_2} \c2 {A_3} \c2
                     {A_4}}
  + \wick{ \c1 {A_1} \c2 {A_2} \c1 {A_3} \c2 {A_4}}
  + \wick{ \c1 {A_1} \c2 {A_2} \c2 {A_3} \c1 {A_4}}.                      
\end{eqnarray}
The tensor decomposition of Eq.~\eqref{eq:TRb} of a generic tensor integral is written compactly as, 
{
\begin{equation}
  \label{eq:genericred_ordering}
  \boxed{
  I^{\mu_1 \ldots \mu_R} =I_{\alpha_1 \ldots \alpha_R} \; 
  \TO{
\prod_{i=1}^R \braket{\alpha_i}{ \mu_i} 
 }.
}
\end{equation}
}
This is the main result of this article.

Let us remark that all multiplications in the terms of the right-hand side of Eq.~\eqref{eq:genericred_ordering} are commutative.  It does not matter, for example, if we have the unit tensors with indices $\alpha_1,\mu_1$ on the right and unit tensors with indices $\alpha_R,\mu_R$ on the left.

\subsection{Rank-2 tensor triangle integrals with two external momenta}
As an instructive example, we demonstrate the steps which lead to Eq.~\eqref{eq:genericred_ordering} in the case of a rank-2 tensor integral $I^{\mu\nu}$ with two independent external legs $p_1,p_2$. 
Our basis of independent tensors consists of
\begin{align}
    \mathcal{B} = \left\{ p_1^{\mu}p_1^{\nu},\,p_1^{\mu}p_2^{\nu},\,p_2^{\mu}p_1^{\nu},\,p_2^{\mu}p_2^{\nu},\,\eta_\perp^{\mu\nu} \right\}.
\end{align}
Correspondingly, the dual basis is given by
\begin{align}
    \vev{\mathcal{B}} = \left\{ \vev{p_1}^{\mu}\vev{p_1}^{\nu},\vev{p_1}^{\mu}\vev{p_2}^{\nu},\vev{p_2}^{\mu}\vev{p_1}^{\nu},\vev{p_2}^{\mu}\vev{p_2}^{\nu},\vev{\eta_\perp^{\mu\nu}} \right\}.
\end{align}
Following \eqref{eq:TRb}, the tensor integral takes the form
\begin{align}
    I^{\mu\nu} &= I_{\alpha\beta} 
    \left[ p_1^{\alpha}p_1^{\beta} \vev{p_1}^{\mu}\vev{p_1}^{\nu}
    + p_1^{\alpha}p_2^{\beta} \vev{p_1}^{\mu}\vev{p_2}^{\nu} 
    + p_2^{\alpha}p_1^{\beta} \vev{p_2}^{\mu}\vev{p_1}^{\nu} \nonumber \right.\\
    &\hspace{4em}\left.+ p_2^{\alpha}p_2^{\beta} \vev{p_2}^{\mu}\vev{p_2}^{\nu} 
    + \eta_\perp^{\alpha\beta} \vev{\eta_\perp^{\mu\nu}} \right] \, . 
\end{align}
Collecting the momentum-dependent tensors with indices $\alpha$ and $\mu$ as well as $\beta$ and $\nu$ allows us to rewrite the expression compactly as
\begin{align}
    I^{\mu\nu} &= I_{\alpha\beta} 
    \left[\left(\sum_{i=1}^2 p_i^{\alpha} \vev{p_i}^{\mu} \right) \left(\sum_{j=1}^2 p_j^{\beta} \vev{p_j}^{\nu}\right) + \eta_\perp^{\alpha\beta} \vev{\eta_\perp^{\mu\nu}} \right].
\end{align}
We can now recognise the unit tensors $u^{\alpha\mu}$ and $u^{\beta\nu}$, defined in Eq.~\eqref{eq:bracket}, for $N_{\text{p}}=2$ external momenta and write
\begin{align}
    I^{\mu\nu} &= I_{\alpha\beta} 
    \left[\braket{\alpha}{\mu}\braket{\beta}{\nu} + \eta_\perp^{\alpha\beta} \vev{\eta_\perp^{\mu\nu}} \right].
\end{align}
The last term corresponds to a contraction of two unit tensors
\begin{eqnarray}
  \label{eq:contraction}
  \wick{\c {\braket{\alpha}{ \mu}} \c {\braket{\beta}{ \nu}}}
  \equiv
\eta_\perp^{\alpha \beta} \, \vev{\eta_\perp^{\mu
  \nu}}
  =\frac{1}{D_\perp} \, \eta_\perp^{\alpha \beta} \, \eta_\perp^{\mu
  \nu}  \, ,
\end{eqnarray}
where, in this example, we have $D_\perp = D-2$.
The tensor integral can be written as
\begin{align}
    I^{\mu\nu} &= I_{\alpha\beta} 
    \left[\braket{\alpha}{\mu}\braket{\beta}{\nu} +  \wick{\c {\braket{\alpha}{ \mu}} \c {\braket{\beta}{ \nu}}} \right],
\end{align}
which resembles the operation of time-ordering and Wick contractions. Using the ordering symbol defined above, we write 
\begin{align}
    I^{\mu\nu} &= I_{\alpha\beta} 
    {\cal T} \left\{ \braket{\alpha}{\mu}\braket{\beta}{\nu} \right\}. 
\end{align}

\section{Illustrative applications}
\label{sec:applications}

Eq.~\eqref{eq:genericred_ordering}  provides an algorithmic instruction for projecting tensor integrals to scalar integrals, in which no large matrix inversions are needed in intermediate steps. 
Eq.~\eqref{eq:genericred_ordering} is a very compact expression, which includes all terms which emerge in a tensor reduction.  
It is useful to be able to just write down such sizable results directly, without intermittent algebraic operations. 
As already mentioned in Section \ref{sec:problem_TR}, the number of possible tensors $N_{\rm tensors}$,  enumerated in Eq.~\eqref{eq:Ntensors},  rises steeply with the rank of the tensor and the number of external particles.

Still, the computational cost of merely casting the terms on the right-hand side of  Eq.~\eqref{eq:genericred_ordering} is substantial. It is further increased if explicit substitutions for dual products of transverse metric tensors, for the transverse metric in terms of the D-dimensional metric and the dual momenta in terms of the original momenta are carried out analytically. 

However, in realistic Feynman diagram computations, many explicit substitutions of the transverse metric $\eta_\perp^{\mu\nu}$ in the elements of the dual basis may not be necessary.  For instance, one can use directly that $\eta_\perp^{\mu\nu} p_{i\nu}=0.$ When the transverse metric is contracted to a gamma matrix, it projects out its transverse components 
\begin{equation}
\label{eq:gammaperp}
    \gamma_\perp^\mu \equiv \eta_\perp^{\mu\nu} \gamma_\nu. 
\end{equation}
We can also simplify spin-chains making use of the Clifford algebra for the transverse components, which in conventional dimensional regularisation reads~\cite{Smith:2004ck}
\begin{equation}
\label{eq:Cliffordperp}
    \left\{ \gamma_\perp^\mu, \gamma^\nu\right\} =  \left\{ \gamma_\perp^\mu, \gamma_\perp^\nu\right\} =  2 \eta_\perp^{\mu\nu} {\bf 1}_{4\times 4}
\end{equation}
and 
\begin{equation}
\label{eq:gammacontractionperp}
\gamma^\mu_\perp \gamma_\mu = \gamma^\mu_\perp \gamma_{\perp, \mu} =  D_\perp \, {\bf 1}_{4\times 4} \, . 
\end{equation}
Also, one can make use of the property of the unit tensor, $p_{i,\mu} \braket{\mu}{\nu} = p_i^\nu $, without expanding $\braket{\mu}{\nu}$ in terms of 
momentum vectors. 

To illustrate how Eq.~\eqref{eq:genericred_ordering} may be used and to assess its practical potential, we implemented the reduction for  tensor integrals with three independent light-like external momenta. We then applied the reduction to the analytic computation of selected Feynman diagrams in  $2\to2 $ scattering amplitudes.  We present these applications next.

\subsection{Tensor reduction for integrals with three independent light-like momenta}

We consider tensor integrals of generic rank $R$
\begin{eqnarray}
I^{\mu_1 \ldots \mu_R}\left( p_1, p_2, p_3 \right),
\end{eqnarray}
which depend on three independent light-like external momenta. 
We define the corresponding Mandelstam variables as
\begin{align}
\label{eq:kinem2to2}
    p_1^2 =p_2^2 = p_3^2 = 0 \, , \; 
    2 \, p_1\cdot p_2 = s \, , \;
    2 \, p_2\cdot p_3 = -t \, ,\, 
    2 \, p_1\cdot p_3 = {s+t} \, .
\end{align}
The matrix of scalar products of the external momenta in our basis, $\Pi_{ij} = p_i \cdot p_j $, can be read off from  Eq.~\eqref{eq:kinem2to2},
\begin{align}
\label{eq:PImatrix}
\Pi  = \frac{1}{2} \left(
\begin{array}{ccc}
  0 & s & s+t \\
  s & 0 & -t \\
  s+t & -t & 0 \\ 
\end{array}
  \right).
  \end{align}
   The inverse matrix can be easily calculated 
  \begin{align}
  \label{eq:PIinvmatrix}
  \Delta = 
  \left(\begin{array}{ccc}
  \frac{t}{s(s+t)} & \frac{1}{s} & \frac{1}{s+t} \\
  \frac{1}{s} & \frac{s+t}{s\,t} & \frac{1}{-t} \\
  \frac{1}{s+t} & \frac{1}{-t} & \frac{s}{t(s+t)}\\ 
\end{array} \right).
\end{align}
The  calculation of the inverse matrix $\Delta$ is the only process-dependent inversion which we require. Its matrix elements are the scalar products of the dual momentum vectors,  $\Delta_{ij} = \vev{p_i} \cdot \vev{p_j}$.  
The dual momenta, $\vev{p_i}, \; i=1..3$, can be computed from the matrix $\Delta$ and  the definition $\vev{p_i} = \sum_{j=1}^3 \Delta_{ij} \, p_j$. 
Finally, we construct the unit tensor $u^{\alpha \mu}$ out of the independent external momenta $p_1,p_2,p_3$, 
\begin{equation}
\label{eq:unittensex}
    \braket{\alpha}{\nu} = \sum_{i=1}^3 p_i^\alpha \vev{p_i}^\nu \, . 
\end{equation}
This is the main building block for creating $ \TO{\prod_{i=1}^R \braket{\alpha_i}{ \mu_i} }$ on the right-hand side of Eq.~\eqref{eq:genericred_ordering}.  

 In the space of tensors which can be constructed out of metric and momenta products,  the $ \TO{\prod_{i=1}^R \braket{\alpha_i}{ \mu_i} }$
acts as a unity.  
We have written  computer programmes which generate $ \TO{\prod_{i=1}^R \braket{\alpha_i}{ \mu_i}}$ for a given rank $R$ and set of independent external momenta.  With explicit computations in \textsc{Form} we have verified that indeed 
\begin{eqnarray}
&& \left( 
\TO{\prod_{i=1}^R \braket{\alpha_i}{ \mu_i} } -
\prod_{i=1}^R \eta^{\alpha_i \mu_i}
\right) 
\, 
 \, { T}_{a_1, \ldots a_R}\left(\eta , {p_i} \right)
=0, 
\end{eqnarray}
for three independent external momenta and tensors up to rank five.

\subsection{One-loop $q \bar q \to \gamma \gamma$ amplitude}
We will now apply our tensor reduction on Feynman diagrams in the scattering amplitude for quark-antiquark annihilation to a pair of real photons, 
\[ 
q(p_1) + \bar q (p_2) \to \gamma(p_3) + \gamma(p_4) \, . 
\]
In parentheses we denote the momenta of the external particles, which satisfy the momentum conservation condition
\begin{equation}
p_1 + p_2  = p_3 +p_4
\end{equation}
and Mandelstam variables as given in Eq.~\eqref{eq:PImatrix}.
All tensor integrals can be reduced to scalar integrals with the expressions for  $\TO{\prod_{i=1}^R \braket{\alpha_i}{ \mu_i} }$ that we have built above.

We start by computing the hard scattering contribution to the one-loop amplitude. We have purposefully chosen to calculate a physical amplitude, comprising of various Feynman diagrams and counterterms, in order to show the versatility of the approach in dealing with diverse  diagrammatic topologies (tadpole, bubble, triangle and box in this particular example).

Following Ref.~\cite{Anastasiou:2022eym}, the process-specific finite amplitude remainder may be defined as 
\begin{equation}
\label{eq:1loopfinite}
\mathcal{M}_{q \bar q \to  \gamma \gamma,\text{finite}}^{(1)} = {\cal H}^{(1)}\left(\{p_3, \epsilon_3 \} , \{p_4, \epsilon_4 \}\right)
+
{\cal H}^{(1)}\left(\{p_4, \epsilon_4 \} , \{p_3, \epsilon_3 \}\right),
\, 
\end{equation}
where 
\begin{eqnarray}
\label{eq:1loop}
    {\cal H}^{(1)}\left(\{p_3, \epsilon_3 \} , \{p_4, \epsilon_4 \}\right)  &=& 
    \includegraphics[width=0.2\textwidth, page=1,valign=c]{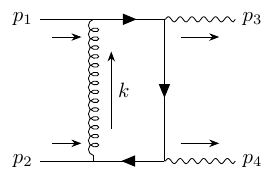} 
    + \, \left( 1 - \mathcal{R}_{k} \right) \; \includegraphics[width=0.2\textwidth, page=2,valign=c]{tikz_TR1loop.pdf}  
           \nonumber \\ && \hspace{-20mm}
    + \, \left( 1 - \mathcal{R}_{k} \right) \;
\includegraphics[width=0.2\textwidth, page=3,valign=c]{tikz_TR1loop.pdf}
     + \, \left( 1 - \mathcal{R}_{k} \right) \;
\includegraphics[width=0.18\textwidth, page=4,valign=c]{tikz_TR1loop.pdf}
       \nonumber \\ && \hspace{5mm}
     - \, \left( 1 - \mathcal{R}_{k} \right) \;
\includegraphics[width=0.23\textwidth, page=5,valign=c]{tikz_TR1loop.pdf}.
\end{eqnarray}
The last diagram is a form-factor counterterm and removes the infrared singularities of the amplitude. The form-factor counterterm is obtained from the pictured Feynmnan diagram by enclosing the part of the diagram in parentheses in between  spin projector factors given by Eq.~(2.4) of Ref.~\cite{Anastasiou:2022eym} and evaluating the corresponding expression at zero gluon momentum $k=0$. Diagrams with one-loop vertex and propagator corrections have ultraviolet singularities.  They are  subjected to an ultraviolet subtraction indicated by $\mathcal{R}_{k}$.  The explicit form of the ultraviolet counterterms, is given by Eq.~(6.7) and Eq.~(6.15) of Ref.~\cite{Anastasiou:2022eym}. 

We examine first  the tensor reduction steps of the box diagram which yields tensors of the highest rank three. The diagram is written as,
\begin{align}
\label{eq:boxdiagram}
&\includegraphics[width=0.25\textwidth, page=1,valign=c]{tikz_TR1loop.pdf}\\
& = -g_s^2Q_q^2e^2\,C_F \int \frac{\mathrm{d}k^D}{(2\pi)^D}
\frac{\bar v (p_2) \gamma^\mu \gamma_{\nu} \slashed \epsilon_4^* \gamma_{\rho} \slashed \epsilon^*_3  \gamma_{\sigma} \gamma_\mu u(p_1)}{k^2 k_1^2 k_2^2 k_3^2  } \, k_1^\nu k_2^\rho k_3^\sigma 
\nonumber \\ 
\hspace{0cm}
&=
 -g_s^2Q_q^2e^2\,C_F \int \frac{\mathrm{d}k^D}{(2\pi)^D}
\frac{\bar v (p_2) \gamma^\mu \gamma_{\nu} \slashed \epsilon^*_4 \gamma_{\rho} \slashed \epsilon^*_3  \gamma_{\sigma} \gamma_\mu u(p_1)}{k^2 k_1^2 k_2^2 k_3^2  } \, {k_1}_\alpha {k_2}_\beta {k_3}_\gamma
\TO{
\braket{\alpha}{ \nu}
\braket{\beta}{ \rho}
\braket{\gamma}{ \sigma}
 }
\end{align}
with 
\begin{equation}
k_1 = k-p_2, \; k_2= k+p_1-p_3, \; k_3 = k+p_1 \, .  
\end{equation}
In the second line of Eq.~\eqref{eq:boxdiagram} we have already inserted the ordering operator of Eq.~\eqref{eq:genericred_ordering} which will project the loop momentum tensors to scalars. 

We now carry out the contractions as instructed by the ordering of the unit tensors, derived in the previous section, in the integrand. The numerator of the diagram reads 
\begin{eqnarray}
\label{eq:TOexp}
&& {k_1}_\alpha {k_2}_\beta {k_3}_\gamma
\TO{
\braket{\alpha}{ \nu}
\braket{\beta}{ \rho}
\braket{\gamma}{ \sigma}
 } =\sum_{i_1 i_2 i_3} (k_1 \cdot p_{i_1})\,(k_2 \cdot p_{i_2}) \, (k_3 \cdot p_{i_3}) \,  
 \vev{p_{i_1}}^\nu 
 \vev{p_{i_2}}^\rho
 \vev{p_{i_3}}^\sigma 
 \nonumber \\ 
 && 
 +  \frac{k_\perp^2}{D-3}    
\sum_{j}   \left[ 
 (k_3 \cdot p_j) \, \eta_{\perp}^{\nu \rho} \, \vev{p_j}^{\sigma}
+ (k_2 \cdot p_j) \, \eta_{\perp}^{\sigma \nu} \,  \vev{p_j}^{\rho}
+ (k_1 \cdot p_j) \, \eta_{\perp}^{\rho \sigma} \,  \vev{p_j}^{\nu}
 \right] \, , 
\end{eqnarray}
with 
\begin{eqnarray}
   && k_{1,\perp}^2 =k_{2,\perp}^2 =k_{3,\perp}^2 =
   k_\perp^2
   \nonumber \\ &&
   \equiv 
   \eta^{\alpha \beta}_\perp k_\alpha k_\beta = 
   k^2 -\sum_{j_1 j_2} \vev{p_{j_1}} \cdot \vev{p_{j_2}} \, k \cdot p_{j_1} \, k \cdot p_{j_2} \, . 
\end{eqnarray}
At this stage, we have projected the tensor product of loop momenta to the scalar products $k^2, k \cdot p_1, k \cdot p_2, k \cdot p_3\,$. These scalar integrals can be expressed in terms of the loop momentum denominators of the diagram, 
\[ 
2 k \cdot p_2 = k^2 -k_1^2, \; 
2 k \cdot p_1 = k_3^2 - k^2, \; 
2 k \cdot p_3 = k_3^2 -k_2^2 -2 p_1 \cdot p_3 \, . 
\]
The resulting scalar integrals are of the form, 
\begin{eqnarray}
\label{eq:Box1Ltopology}
    {\rm TP}\left( i_1, i_2, i_3, i_4\right)
    =\int \frac{d^Dk}{(2 \pi)^D}  
    \frac{1}{
    \left(k^2\right)^{i_1}
    \, 
    \left(k_3^2\right)^{i_2}
    \, 
    \left(k_2^2\right)^{i_3}
    \,  
    \left(k_1^2\right)^{i_4}
    } \, , 
\end{eqnarray}
with integer powers $i_j \leq 1$. 

We have already reduced the tensor integral to scalar integrals. We can now proceed with further simplifications pertinent to the spin and Lorentz structure of the diagram. 
We substitute 
\begin{equation}
\label{eq:etaperp2to2}
\eta^{\nu \rho}_\perp = 
\eta^{\nu \rho} - \braket{\nu}{\rho} 
= \eta^{\nu \rho} - \sum_{i,j=1}^{3} 
\Delta_{ij} \, 
p_i^\nu  \, p_j^\rho   
\end{equation}
into Eq.~\eqref{eq:TOexp} and express the vectors of the dual momenta $\vev{p_i^\nu}$ 
in the second line of Eq.~\eqref{eq:TOexp} 
as linear combinations of the external momenta
\begin{equation}
    \vev{p_i^\mu} = \sum_{j} 
    \Delta_{ij}
    \, p_j^\mu. 
\end{equation}
The vectors $p_1, p_2, p_3$ and the metric tensor are contracted with gamma matrices. We can simplify the spin chains using the Clifford algebra, in conventional dimensional regularisation, 
and that external states are on-shell, 
\begin{align}
\slashed{p}_1 u(p_1)=0, \; \bar{v}(p_2)\, \slashed{p}_2 = 0,\; \epsilon^*(p_3)\cdot p_3 = 0, \; \; \epsilon^*(p_4)\cdot p_4 = 0.
\end{align}
The polarisation vectors are transverse to reference vectors that we choose as 
\begin{align}
    \epsilon^*(p_3)\cdot p_4 = 0, \; \; \epsilon^*(p_4)\cdot p_3 = 0.
\label{eq:polchoice}
\end{align}
After these simplifications, the diagram is expressed in terms of the following minimal set of spin-chains,  
\begin{eqnarray}
\label{eq:minsetspinchain}
 {S}_1   &=&\bar{v}(p_2) \slashed{\epsilon}^*_4(\slashed{p}_1-\slashed{p}_3) \slashed{\epsilon}^*_3 u(p_1),
    \nonumber \\ 
   S_2 &=& 
    \bar{v}(p_2)\slashed{p}_3u(p_1), \; 
   \nonumber \\
S_3 &=&     
\bar{v}(p_2)\slashed{\epsilon}^*_3u(p_1),\; 
\nonumber \\ 
S_4 &=&
\bar{v}(p_2)\slashed{\epsilon}^*_4u(p_1) \, .
\end{eqnarray}

Our treatment of the other Feynman diagrams is analogous. Although the loop denominators for triangle and bubble graphs depend on just two or one combinations of the $p_1,p_2,p_3$ external momenta, we can decide to treat all diagrams uniformly.
For example, the tensor loop integral in the second Feynman diagram of Eq.~\eqref{eq:1loop}, which is a triangle, depends only on the 
\[ 
q_a=p_2,\; q_b=p_1+p_2-p_3, 
\]
combinations of external momenta. We could use ${\cal B}_2=\{q_a,q_b\}$ as our basis of vectors for the tensor reduction of this particular diagram. However, it is also allowed and, perhaps, preferred to use the bigger basis ${\cal B}_{3}=\{p_1, p_2, p_3\}$ which we needed for the box diagram earlier. The only disadvantage of using the extended basis ${\cal B}_3$ is that it results to scalar integrals 
${\rm TP}(i_1, i_2, i_3, i_4)$ with some negative powers of propagators $i_j$. However, scalar integrals with negative propagator powers can be handled with integration by parts identities and the Laporta algorithm~\cite{Laporta:2001dd}  seamlessly. 
In addition, beyond one loop, negative powers of propagators (or, equivalently, irreducible scalar products in the numerators) are inevitable. 

We only treat ultraviolet counterterms, denoted by $R_k$ in Eq.~\eqref{eq:1loop}, separately. These give rise to second rank tensor integrals of a tadpole topology. It is very easy to reduce tensors of the tadpole topology in their natural basis, consisting of metrics and no external momenta. We write  
\begin{eqnarray}
\label{eq:tadpoletensor}
&& \int \frac{d^D k}{(2 \pi)^D} 
\frac{
k^{\mu_1} k^{\mu}
}{\left(k^2- M^2 \right)^n} 
=  
\vev{\eta^{\mu_1 \mu_2}} \, \int \frac{d^D k}{(2 \pi)^D} 
\frac{
k^2 
}{\left(k^2- M^2 \right)^n}  \, . 
\end{eqnarray}
In the above, the dual of the metric is $\vev{\eta^{\mu_1 \mu_2}} = \frac{\eta^{\mu_1 \mu_2}}{D}$. 

At this point, we have achieved our goal of reducing all tensor integrals in the amplitude to scalar integrals.  For a complete analytic computation of the amplitude, one may further reduce these scalar integrals to master integrals. 
We reduce all scalar integrals $ \mathrm{TP}(i_1, i_2, i_3, i_4)$  to the one-loop box and the one-loop bubble master integrals with integration by parts identities using \textsc{AIR}~\cite{Anastasiou:2004vj}, 
\begin{eqnarray}
\mathrm{TP}(i_1, i_2, i_3, i_4) 
&=& c_4(i_1, i_2, i_3, i_4) \, \, 
\mathrm{TP}(1, 1, 1, 1)
+c_{2a}(i_1, i_2, i_3, i_4) \, 
\mathrm{TP}(1, 0, 1, 0)
\nonumber \\
&& +c_{2b}(i_1, i_2, i_3, i_4) \, 
\mathrm{TP}(0, 1, 0, 1). 
\end{eqnarray}
Inserting the known master integrals (see, for example, Appendix C of Ref~\cite{Anastasiou:2002zn}),  we obtain a finite expression in $D=4$ dimensions as it is anticipated. 
We find 
\begin{align}
  {\cal H}^{(1)}\left(\{p_3, \epsilon_3 \} , \{p_4, \epsilon_4 \}\right)
  &= -i  \frac{Q_q^2 e^2 \alpha_s^{0} \, C_F}{4 \pi} \frac{1}{-s-t}\left( 
S_1  h_1^{(1)}(s, t) \right.\nonumber\\
 &\left.+S_2 \left( \epsilon^*_3\cdot\epsilon^*_4\, h_2^{(1)}(s, t)
+ \epsilon^*_3\cdot p_1 \,\epsilon^*_4\cdot p_2\,  \frac{1}{s} h_3^{(1)}(s, t)        \right)\right.\nonumber\\
 &\left.+(S_3\, \epsilon^*_4\cdot p_2 -S_4\,\epsilon^*_3\cdot p_1) h_4^{(1)}(s, t) \right)
 \label{eq:1loop_result}
\end{align}
with the coefficients $h_i^{(1)}(s, t)$ displayed in Appendix \ref{sec:Ap_coeff}.
%
As a consistency verification, we interfere the one-loop amplitude with the tree-level and obtain a result independent of our tensor reduction.


\subsubsection{$N_f$ contribution to the two-loop  $q \bar q \to \gamma \gamma$ amplitude}

Our tensor reduction depends only on the rank of tensors and the external momenta of Feynman diagrams. Thus, we can use the same reduction expressions  across loop orders, for tensors integrals with common external momenta and rank. For example, it is  very easy to extend the previous one-loop amplitude computation, and  derive similarly the $N_f$ contribution to the finite part of the two-loop $q \bar q \to \gamma \gamma$ amplitude. 

\begin{figure}[H]
    \centering
    \includegraphics[width= 0.28\textwidth, page=6,valign=c]{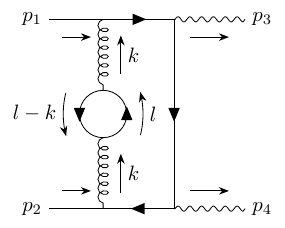}
    \caption{$N_f$ box diagram.}
    \label{fig:BoxNf}
\end{figure}

We generate the $N_f$ part of the two-loop amplitude from the corresponding one-loop amplitude, by inserting a one-loop fermion loop in the gluon propagator. As an example, the box diagram in the one-loop amplitude in Eq.~\eqref{eq:1loop} with a quark self-energy inserted in the gluon propagator is displayed in Fig.~\ref{fig:BoxNf}.

We will consider the case of massless quarks in the fermion loop. As discussed in Ref.~\cite{Anastasiou:2020sdt}, we can simplify the integrand  by a first tensor reduction of the fermion-loop subgraph (reducing tensors in the $l$ momentum integration)  and an elimination of terms in the integrand which cancel, due to gauge invariance, in the sum of diagrams.  
The integrand for the finite part of the amplitude then reads,  
\begin{equation}
\mathcal{M}^{(2,N_f)}_{\,q \bar q \to  \gamma \gamma,\text{finite}} = {\cal H}^{(2,N_f)}\left(\{p_3, \epsilon_3 \} , \{p_4, \epsilon_4 \}\right)
+
{\cal H}^{(2,N_f)}\left(\{p_4, \epsilon_4 \} , \{p_3, \epsilon_3 \}\right),
\end{equation}
with 
\begin{align}
    {\cal H}^{(2,N_f)}\left(\{p_3, \epsilon_3 \} , \{p_4, \epsilon_4 \}\right) &= i\, g_s^2\,T_F\, N_f \frac{2(D-2)}{D-1} \left( \frac{1}{l^2(l-k)^2} - \frac{1}{(l^2 -M^2)^2}\right)\,\nonumber\\
    &\quad \times {\cal H}^{(1)}\left(\{p_3, \epsilon_3 \} , \{p_4, \epsilon_4 \}\right).
\end{align}
We now proceed to the tensor reduction of the $k$-momentum integrals. 
The ${\cal H}^{(2,N_f)}\left(\{p_3, \epsilon_3 \}\right)$  has an identical tensor numerator as ${\cal H}^{(1)}\left(\{p_3, \epsilon_3 \}\right)$. Hence we reduce the tensor  in ${\cal H}^{(2,N_f)}$  to scalar products exactly as in ${\cal H}^{(1)}$. 

After the reduction to scalar integrals, we perform a further reduction to two-loop master integrals solving integration by parts identities, using \textsc{AIR}. The two-loop master integrals are known analytically for $2 \to 2$ massless QCD scattering processes~\cite{Tausk:1999vh,Smirnov:1999gc,Anastasiou:2000kp}. We have taken the master integral  expressions from a computer readable input used in Ref.~\cite{Anastasiou:2002zn}.   
After all substitutions, we arrive at the following result for the finite $N_f$ two-loop  $q \bar q \to \gamma \gamma$ amplitude, 
\begin{align}
    {\cal H}^{(2,N_f)}\left(\{p_3, \epsilon_3 \} , \{p_4, \epsilon_4 \}\right)
&= \frac{4\,i}{3} \frac{Q_q^2e^2(\alpha_s^0)^2 N_f T_F C_F}{(4\pi)^2} \frac{1}{-s-t} \left( 
S_1  h_1^{(2,N_f)}(s, t) \right.\nonumber\\
 &\left.+S_2 \left( \epsilon^*_3\cdot\epsilon^*_4\, h_2^{(2,N_f)}(s, t)
+ \epsilon^*_3\cdot p_1 \,\epsilon^*_4\cdot p_2\,  \frac{1}{s} h_3^{(2,N_f)}(s, t)        \right)\right.\nonumber\\
 &\left.+(S_3\, \epsilon^*_4\cdot p_2 - S_4\,\epsilon^*_3\cdot p_1) h_4^{(2,N_f)}(s, t) \right) \, . 
 \label{eq:Nf_result}
\end{align}
The coefficients  $h_i^{(2,N_f,j)}(s, t)$ are  displayed in Appendix \ref{sec:Ap_coeff} and they are free of $1/\epsilon$ poles, as it is anticipated.

\subsection{A two-loop diagram with rank-$5$ tensors in dimensional regularisation} 

The previous illustrative examples are computationally simple, as they require tensor integrals of a relatively low rank, i.e. three. 
We would like to test the implementation of Eq.~\eqref{eq:genericred_ordering} on Feynman integrals with higher rank tensors. 
We find rank-5 tensors (the maximum rank for $2 \to 2$ QCD scattering at the two-loop order) in  Feynman diagrams for the $q \bar q \to \gamma \gamma$ amplitude with seven propagators. As a representative case, we examine a Feynman diagram of a planar double-box topology together with a suitable form-factor type of counterterm~\cite{Anastasiou:2022eym,Anastasiou:2020sdt,Anastasiou:2018rib} which removes a double soft singularity as $k,l \to 0$, 
\begin{align}
\label{eq:Planar2Ldiagram}
    & \mathcal{D}_{2}^{(2)} = \includegraphics[width=0.325\textwidth, page=1,valign=c]{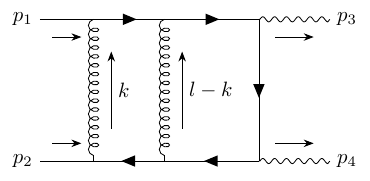} - \includegraphics[width=0.4\textwidth, page=2,valign=c]{tikz_TR2loop.pdf}.
\end{align}
The double-box diagram, first term on the right-hand side of Eq.~\eqref{eq:Planar2Ldiagram}, yields an $1/\epsilon^4$ pole. This pole is cancelled against the contribution of the form factor counterterm, which is the second term on the right-hand side of Eq.~\eqref{eq:Planar2Ldiagram}. This counterterm is again obtained by enclosing the part of the diagram in parentheses in between spin projectors and evaluating the enclosed expression at zero momenta $k,l=0$. We therefore anticipate  that the combination of Eq.~\eqref{eq:Planar2Ldiagram}  has a Laurent series expansion starting at the $1/\epsilon^3$ order. This will be a test for the correctness of our tensor reduction in this example.

The two diagrams consist of tensor integrals of rank $\leq 5$  and depend on three independent light-like momenta, $p_1, p_2, p_3$. The Mandelstam variables have been defined in Eq. \eqref{eq:kinem2to2} and the matrix of scalar products as well as the inverse have been defined 
in Eq.~\eqref{eq:PImatrix} and Eq.~\eqref{eq:PIinvmatrix} respectively. We again use the choice of polarisation vectors defined in Eq.~\eqref{eq:polchoice}. We examine, first,  the double-box diagram which yields tensors of the highest rank-$5$ tensor structures more carefully. The diagram is given by  
\begin{align}
&\includegraphics[width=0.325\textwidth, page=1,valign=c]{tikz_TR2loop.pdf}\nonumber\\
& = ig_s^4Q_q^2e^2C_F^2  \int \frac{\mathrm{d}k^D}{(2\pi)^D}\frac{\mathrm{d}l^D}{(2\pi)^D}
\frac{\bar v (p_2) \gamma^{\nu} \gamma_{\mu_1}  \gamma^{\sigma} \gamma_{\mu_2} \slashed \epsilon^*_4 \gamma_{\mu_3} \slashed \epsilon^*_3  \gamma_{\mu_4} \gamma_{\sigma} \gamma_{\mu_5} \gamma_{\nu} u(p_1)}{k^2\,(l-k)^2\, k_1^2\, k_2^2\, k_3^2\, k_4^2\, k_5^2  } \, k_1^{\mu_1}\,  k_2^{\mu_2}\, k_3^{\mu_3}\, k_4^{\mu_4}\, k_5^{\mu_5}
\nonumber \\ 
\hspace{0cm}
&= ig_s^4Q_q^2e^2C_F^2  \int \frac{\mathrm{d}k^D}{(2\pi)^D}\frac{\mathrm{d}l^D}{(2\pi)^D}
\frac{\bar v (p_2) \gamma^{\nu} \gamma_{\mu_1}  \gamma^{\sigma} \gamma_{\mu_2} \slashed \epsilon^*_4 \gamma_{\mu_3} \slashed \epsilon^*_3  \gamma_{\mu_4} \gamma_{\sigma} \gamma_{\mu_5} \gamma_{\nu} u(p_1)}{k^2\,(l-k)^2\, k_1^2\, k_2^2\, k_3^2\, k_4^2\, k_5^2  } \, \nonumber \\
& \hspace{2cm} \times k_{1,\alpha_1}\,  k_{2,\alpha_2}\, k_{3,\alpha_3}\, k_{4,\alpha_4}\, k_{5,\alpha_5}
\TO{
\braket{\mu_1}{ \alpha_1}\braket{\mu_2}{ \alpha_2}\braket{\mu_3}{ \alpha_3}\braket{\mu_4}{ \alpha_4}\braket{\mu_5}{ \alpha_5} },
    \label{eq:Dboxdiagram}
\end{align}
where
\begin{align}
    k_1 = k-p_2,\; k_2 = l-p_2,\; k_3 = l+p_1-p_3,\; k_4 = l+p_1, \, k_5 = k+p_1.
\end{align}
In the third line of Eq.~\eqref{eq:Dboxdiagram} we 
included the ordering operator defined in Eq.~\eqref{eq:genericred_ordering}, which projects the tensor integral to scalar integrals. Note that we have switched the indices in the unit tensors $u^{\mu\alpha}$ enclosed by the ordering operator. In contrast to the one-loop case, we have chosen here to contract the loop momentum tensors with the dual metric tensors and dual external momentum vectors, whereas the non-dual transverse metric and momentum tensors are getting contracted with the gamma matrices in the spin-chains. This choice corresponds to the reduction in Eq.~\eqref{eq:TRa} instead of the previously used reduction in Eq.~\eqref{eq:TRb}. 
Our choice is motivated by computational optimisation. Now, we can proceed directly to  simplifying the spin-chains without substituting the dual tensors in terms of momenta and metrics. 

The ordering of the five unit tensors in Eq.~\eqref{eq:Dboxdiagram} leads to several terms such as a product of five unit tensors, a product of three unit tensors times one transverse metric and its dual, and so on. To enhance the performance of the code, we apply an iterative procedure to reduce the spin-chains. We explicitly substitute only one unit tensors in terms of the momenta and their dual momenta defined in Eq.~\eqref{eq:unittensex} at a time. The external momenta from the unit tensor contracts with the gamma matrices in the spin-chains. This allows us to simplify the spin-chains using on-shellness of the external momenta and the Clifford algebra. We iterate this procedure, replacing one unit tensor at a time and simplifying the spin-chains, until there are no unit tensors left. This leaves the contraction of gamma matrices with the transverse metric tensors, which projects out the transverse component $\gamma_{\perp}$ defined in Eq.~\eqref{eq:gammaperp}. We can make use of the transverse Clifford algebra in Eq.~\eqref{eq:Cliffordperp}
and~\eqref{eq:gammacontractionperp}. These simplifications lead again to the minimal set of spin-chains displayed in Eq.~\eqref{eq:minsetspinchain}.

As a last step we have to reduce the loop momenta contracted with dual tensors to scalar integrals such that they allow for reduction to master integrals with integration by parts methods. The three scalar product structures we encounter after the insertion of the ordering operator are given by
\begin{align}
    &k_{1\alpha_1}k_{2\alpha_2} k_{3\alpha_3}k_{4\alpha_4} \vev{\eta_{\perp}^{\alpha_1\alpha_2}\eta_{\perp}^{\alpha_3\alpha_4}} k_5 \cdot \vev{p_1},\nonumber\\
    &k_{1\alpha_1}k_{2\alpha_2}  \vev{\eta_{\perp}^{\alpha_1\alpha_2}} k_{3}\cdot\vev{p_1}\, k_{4}\cdot\vev{p_2}\,k_5 \cdot \vev{p_3}, \nonumber\\
    &k_{1}\cdot\vev{p_2}\, k_{2}\cdot\vev{p_1}\, k_{3}\cdot\vev{p_1}\, k_{4}\cdot\vev{p_2}\,k_5 \cdot \vev{p_3},
\end{align}
where we chose some random indices as examples. Since the metric duals are transverse to the external momenta, we can immediately rewrite the contractions with $k_i$, only in terms of the loop momenta. We can simplify the three structures even further by invoking the dual property between the external momenta and their corresponding duals,
as seen in Eq.~\eqref{eq:mom_dual_prop}. 
The three scalar product structures therefore simplify to
\begin{align}
    &k_{\alpha_1}l_{\alpha_2} l_{\alpha_3}l_{\alpha_4} \vev{\eta_{\perp}^{\alpha_1\alpha_2}\eta_{\perp}^{\alpha_3\alpha_4}} (k \cdot \vev{p_1} +1),\nonumber\\
    &k_{\alpha_1}l_{\alpha_2}  \vev{\eta_{\perp}^{\alpha_1\alpha_2}} (l\cdot\vev{p_1}+1)\, l\cdot\vev{p_2}\,k\cdot \vev{p_3}, \nonumber\\
    &(k\cdot\vev{p_2}-1)\, l\cdot\vev{p_1}\, (l\cdot\vev{p_1}+1)\, l\cdot\vev{p_2}\,k\cdot \vev{p_3}.
\end{align}
Additionally, we include the definition of the dual transverse metrics defined in Eq.~\eqref{eq:dualmetricR2} for rank 2 and Eq.~\eqref{eq:dualmetricR4} for rank 4 and substitute the definition for the transverse metric displayed in Eq.~\eqref{eq:etaperp2to2}. As a last step the dual momenta are substituted in terms of external momenta, completing the reduction of the tensor integral to scalar integrals including the scalar products 
\begin{align}
    k^2, \, k\cdot p_1, \, k\cdot p_2, \, k\cdot p_3,\, l^2, \, l\cdot p_1, \, l\cdot p_2, \, l\cdot p_3,\, k\cdot l.
\end{align}

In a similar fashion as in the one-loop case, the obtained scalar integrals can be reduced to master integrals with integration by parts using \textsc{AIR}~\cite{Anastasiou:2004vj}. 

The treatment of the counterterm diagram is analogous, where we keep all three external momenta $p_1,\,p_2,\,p_3$ as a basis for the tensor reduction. Expanding the master integrals in $\epsilon$, the sum of diagrams is given by
\begin{align}
    \mathcal{D}_{2,0,0}^{(2)} = \frac{-i}{s+t}\frac{\left(\alpha_s^{0}\right)^2\,e^2\,Q_q^2\,C_F^2}{\left(4\pi\right)^2}\left(\left( 2\, S_1 + 4\,\epsilon^*_4\cdot p_2 \, S_3 - 4\,\epsilon^*_3\cdot p_1 \, S_4 \right)\times \frac{1}{\varepsilon^3} + \mathcal{O}(\varepsilon^{-2})\right),
\end{align}
where we have used the minimal set of spin-chains defined in Eq.~\eqref{eq:minsetspinchain}. We note that the $1/\varepsilon^4$ pole cancels as expected.
Additionally, we have verified the independence on tensor reduction 
in the interference with the tree-level, as in the one-loop case.



\subsection{A three-loop diagram with rank-7 tensors in dimensional regularisation} 
As a last example, we apply this tensor reduction approach to one three-loop diagram  for the same process ($q\bar q \to \gamma\gamma$) considered above
\begin{align}
    {\cal D}_3^{(3)}
    = \includegraphics[width=0.425\textwidth, page=1,valign=c]{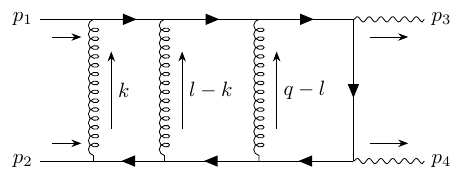}.
\end{align}
This diagram is a rank-7 tensor integral and has the same kinematic structure as the examples discussed above and the same choice of polarisation vectors defined in Eq.~\eqref{eq:polchoice}. We reduce the tensor structure analogously as in Eq.~\eqref{eq:Dboxdiagram}. Hence, we include the ordering operator such that the loop momenta contract with the dual momenta and metrics, whereas the non-dual momenta and metric tensors contract with gamma matrices in the spin-chain function. We take the same steps to simplify the spin-chain down to a minimal set of spin-chains seen in Eq.~\eqref{eq:minsetspinchain}. We leave the scalar integrals in terms of the scalar products
\begin{align}
    &k^2, \, k\cdot p_1, \, k\cdot p_2, \, k\cdot p_3,\nonumber\\
    & l^2, \, l\cdot p_1, \, l\cdot p_2, \, l\cdot p_3,\nonumber\\
    &q^2, \, q\cdot p_1, \, q\cdot p_2, \, q\cdot p_3,\nonumber\\
    &k\cdot l,\, k\cdot q,\, l\cdot q.
\end{align}
without reducing to master integrals. The tensor reduction results in the following number of scalar integrals
\begin{align}
   \mathcal{D}_3^{(3)} = & g_s^6\,e^2\,Q_q^2\,C_F^3 \nonumber\\ & \times \left( S_1 \times  \text{[1467 scalar integrals]} 
   \;+S_2 \times \text{[2560 scalar integrals]}  \right.\nonumber\\
    &\left. \quad+S_3 \times\text{[2557 scalar integrals]} \; +S_4 \times \text{[2559 scalar integrals]} \right)
\end{align}
and was performed in \textsc{TForm}~\cite{Tentyukov_2010} with 16 workers in $1107.30$ seconds on a common desktop. Finally, we were able to confirm the independence on tensor reduction for the three-loop diagram by interfering with the tree-level, as in the previous cases.

\section{Conclusions}
\label{sec:conclusions}

In Eq.~\eqref{eq:genericred_ordering}, we presented a closed-form solution to the diagonalisation problem of the Passarino-Veltman reduction method for arbitrary tensor integrals. We derived this formula by noticing a connection between the tensor elements appearing in a tensor reduction and the ordering, along with associated Wick contractions, of operators that we assigned to tensor indices.

This algebraic analogy of tensor reduction and ordering becomes a precise equality in Equation~\eqref{eq:genericred_ordering} when we choose an appropriate basis of tensor elements. To achieve this, we adopt a tensor basis composed of independent external momenta and metric tensors transverse to external momenta. Additionally, we construct a second \textit{dual} tensor basis that is orthogonal to the first one, combining the constructions presented in Ref.\cite{vanNeerven:1983vr} for momentum vectors and Ref.\cite{Ruijl:2018poj} for metric tensors.

The tensor reduction technique presented in this paper is independent of the loop order and topology of the integral in the sense that it remains identical for all integrals with the same tensor rank and external legs. The key advantage of our method is that it avoids large matrix inversions, with the only requirement being finding the dual momenta of van Neerven and Vermaseren. This significantly reduces the computational cost of tensor reduction.

Our tensor reduction formula enables the treatment of complex Feynman diagrams, even within conventional dimensional regularisation. Extending the use of Passarino-Veltman reduction, our method becomes competitive in cutting-edge computations. In this work, we demonstrated its potential for QCD calculations by applying the method to diagrams with  up to three loops and four external legs, resulting in tensors of rank seven. Throughout the article, we performed Feynman diagram computations analytically.

We are confident that Equation~\eqref{eq:genericred_ordering} can be further applied in realistic calculations, perhaps in more inventive ways. 
For instance, one could envisage numerical evaluations of the multiplications of elements from the dual basis and Feynman diagrams, while scalar integral coefficients can be determined using independent (analytic or numeric) methods. We eagerly anticipate future applications of our method.


\section*{Acknowledgements}
We are thankful to Federico Buccioni, Valentin Hirschi, Dario Kermanschah, Zoltan Kunszt, Achilleas Lazopoulos, Costas Papadopoulos, Lorenzo Tancredi and  Giulia Zanderighi for useful discussions. We thank Matthias Weisswange for his help to uncover an error in Eq. \eqref{eq:rank8appendix} in the previous version of this paper.

\appendix

\section{Dual transverse metrics of rank six and eight}
\label{sec:Appendix}
We denote the transverse metric product of rank six as
\begin{align}
    \eta_{\perp}^{\mu_1\mu_2} \eta_{\perp}^{\mu_3\mu_4} \eta_{\perp}^{\mu_5\mu_6}
    &\equiv (12)(34)(56).
\end{align}
At rank six we encounter 15 independent transverse metric tensors 
\begin{align}
 \{& (12)(34)(56),\,(12)(35)(46),\,(12)(36)(45),\,(15)(34)(26),\,(16)(34)(25),\nonumber\\
      & (13)(24)(56),\,(14)(23)(56),\, (13)(25)(46),\,(13)(45)(26),\, (14)(25)(36),\,\nonumber\\
      & (14)(35)(26),\,  (15)(23)(46),\,(15)(24)(36),\, (16)(24)(35),\,(16)(23)(45)
    \}\, ,    \label{eq:rank6_indepmetric}
\end{align}
with which we can write an ansatz for the dual transverse metric product 
\begin{align}
   \vev{(12)(34)(56)} \equiv
    \vev{\eta_{\perp}^{\mu_1\mu_2}\eta_{\perp}^{\mu_3\mu_4} \eta_{\perp}^{\mu_5\mu_6}}~.
\end{align}
This tensor has to be invariant under several symmetry transformations such as $1\leftrightarrow 2,\, 3\leftrightarrow 4,\, 5\leftrightarrow 6$ and $(12)\leftrightarrow(34),\,(12)\leftrightarrow(56),\,(34)\leftrightarrow(56)$ as well as combinations thereof.
For example the term $(13)(24)(56)$ is trivially invariant under $5\leftrightarrow 6$ but is not invariant under $1\leftrightarrow 2$ or $3\leftrightarrow 4$. The combination defined as 
\begin{align}
     ( 1\underparenthesis{ 2 )( 3} 4)(56) \equiv (13)(24)(56)+(14)(23)(56)
\end{align}
is invariant under these symmetries. This implies that in the ansatz of the dual transverse metric tensor the coefficients of these two metric tensors need to be equal. This term is still not invariant under a symmetry transformation $(12)\leftrightarrow(56)$ or $(34)\leftrightarrow(56)$. Adding the terms 
$(1\underparenthesis{ 2 )(5}6)( 34)$ with the same coefficients used for $( 1\underparenthesis{ 2 )( 3} 4)(56)$ solves the first issue and still ensures the symmetries addressed above. The fully invariant combination under all symmetry transformations is denoted as
\begin{align}
    [(1\underparenthesis{ 2 )( 3}4)(56)] \equiv (1\underparenthesis{ 2 )( 3}4)(56) +(1\underparenthesis{ 2 )(5}6)( 34)+( 3\underparenthesis{4)(5}6)(1 2 )\,.
\end{align}
It indicates that one transverse metric tensor is the same as the original one for which we are trying to find the dual. Similar considerations can be made for metric tensor where all metric tensors differ from the original one. Only the sum of all these terms remain invariant under the symmetry transformations. We denote this class as
\begin{align}
    [(1\underparenthesis{ 2 )( 3}\underparenthesis{ 4)( 5}6)]  \equiv& (1\underparenthesis{\underparenthesis{ 2 )( 3}4)(5}6) +(1\underparenthesis{ 2 )( 3}\underparenthesis{4)(5}6) \nonumber\\
    =& (13)(25)(46)+(13)(45)(26)+ (14)(25)(36)+ (14)(35)(26) \nonumber\\
    &+  (15)(23)(46)+(15)(24)(36)+ (16)(24)(35)+(16)(23)(45) .   
\end{align}
In terms of the usual notation we can write an ansatz for a dual transverse metric product of rank 6 as 
\begin{align}
    \vev{\eta_{\perp}^{\mu_1\mu_2}\eta_{\perp}^{\mu_3\mu_4} \eta_{\perp}^{\mu_5\mu_6}}
    &= 
    a \; \eta_{\perp}^{\mu_1\mu_2}\eta_{\perp}^{\mu_3\mu_4} \eta_{\perp}^{\mu_5\mu_6}\nonumber\\
    &+ b \,\left( \eta_{\perp}^{\mu_1\mu_3}\eta_{\perp}^{\mu_2\mu_4} \eta_{\perp}^{\mu_5\mu_6} + ...\right)\nonumber\\
    &+ c \,\left( \eta_{\perp}^{\mu_1\mu_3}\eta_{\perp}^{\mu_2\mu_5} \eta_{\perp}^{\mu_4\mu_6} + ...\right),
\end{align}
where each class has a different coefficient. The coefficients are determined using the orthogonality relations between the different metric tensor products of each type, such that we find
\begin{align}
  &D_\perp (D_\perp - 1) (D_\perp - 2) (D_\perp + 4) (D_\perp + 2) \vev{\eta_{\perp}^{\mu_1 \mu_2} \eta_{\perp}^{\mu_3 \mu_4} \eta_{\perp}^{\mu_5 \mu_6}} \nonumber\\
  &= \left( D_\perp^2+3 D_\perp-2\right) \eta_{\perp}^{\mu_1 \mu_2} \eta_{\perp}^{\mu_3 \mu_4}\eta_{\perp}^{\mu_5 \mu_6} \\
  &\hspace{1cm} -\left( D_\perp+2 \right)
  \left(\eta_{\perp}^{\mu_1 \mu_2} \eta_{\perp}^{\mu_3 \mu_5} \eta_{\perp}^{\mu_4 \mu_6}
  +\eta_{\perp}^{\mu_1 \mu_2} \eta_{\perp}^{\mu_3 \mu_6} \eta_{\perp}^{\mu_4 \mu_5}
  +\eta_{\perp}^{\mu_3 \mu_4} \eta_{\perp}^{\mu_1 \mu_5} \eta_{\perp}^{\mu_2 \mu_6}
     \right.
\nonumber  \\
&  \hspace{3.5cm} \left.
   +\eta_{\perp}^{\mu_3 \mu_4} \eta_{\perp}^{\mu_1 \mu_6} \eta_{\perp}^{\mu_2 \mu_5}
   +\eta_{\perp}^{\mu_5 \mu_6} \eta_{\perp}^{\mu_1 \mu_3} \eta_{\perp}^{\mu_2 \mu_4}
   +\eta_{\perp}^{\mu_5 \mu_6} \eta_{\perp}^{\mu_1 \mu_4} \eta_{\perp}^{\mu_2 \mu_3}
   \right)
\nonumber \\
  & \hspace{1cm}
     + 2  \left(
     \eta_{\perp}^{\mu_1 \mu_3} \eta_{\perp}^{\mu_2 \mu_5} \eta_{\perp}^{\mu_4 \mu_6}
     + \eta_{\perp}^{\mu_1 \mu_3} \eta_{\perp}^{\mu_2 \mu_6} \eta_{\perp}^{\mu_4 \mu_5}
     + \eta_{\perp}^{\mu_1 \mu_4} \eta_{\perp}^{\mu_2 \mu_6} \eta_{\perp}^{\mu_3 \mu_5}
     + \eta_{\perp}^{\mu_1 \mu_4} \eta_{\perp}^{\mu_2 \mu_5} \eta_{\perp}^{\mu_3 \mu_6}
     \right.
     \nonumber \\
  & \hspace{2cm}
     \left.
     + \eta_{\perp}^{\mu_1 \mu_5} \eta_{\perp}^{\mu_2 \mu_3} \eta_{\perp}^{\mu_4 \mu_6}
     + \eta_{\perp}^{\mu_1 \mu_5} \eta_{\perp}^{\mu_2 \mu_4} \eta_{\perp}^{\mu_3 \mu_6}
     + \eta_{\perp}^{\mu_1 \mu_6} \eta_{\perp}^{\mu_2 \mu_3} \eta_{\perp}^{\mu_4 \mu_5}
     + \eta_{\perp}^{\mu_1 \mu_6} \eta_{\perp}^{\mu_2 \mu_4} \eta_{\perp}^{\mu_3 \mu_5}
     \right)\nonumber
   \, , 
\end{align}
where we have listed the full expression with every metric tensor. 

Similarly, we can construct the dual for the product of transverse metric tensors of rank eight. The 105 independent transverse metric tensors at rank eight are grouped into five classes w.r.t. exchanges. The class representatives are given by
\begin{align}
     &[(12)(34)(56)(78)],\,\\
     &[(1\underparenthesis{ 2 )( 3}4)(56)(78)],\,\\
     &[(1\underparenthesis{ 2 )( 3}4)(5\underparenthesis{ 6 )( 7}8)]\label{eq:disconnected2},\,\\&[(1\underparenthesis{ 2 )( 3}\underparenthesis{ 4 )( 5}6)(78)],\,
     \\
     &[(1\underparenthesis{ 2 )( 3}\underparenthesis{ 4 )( 5}\underparenthesis{ 6 )( 7}8)]\label{eq:connected4}\,.
\end{align}
Each of these classes are a combination of independent transverse metric products invariant under all symmetry transformations needed for $\vev{\eta_{\perp}^{\mu_1\mu_2}\eta_{\perp}^{\mu_3\mu_4} \eta_{\perp}^{\mu_5\mu_6}\eta_{\perp}^{\mu_7\mu_8}}$. We note that the terms in Eqs.~\eqref{eq:disconnected2} and in \eqref{eq:connected4} both have all pairs of indices different from the original class. Both cases are separately invariant under all possible symmetry transformations.
Hence we will give to each class a different coefficient in the ansatz. In terms of the usual notation we get
\begin{align}
    \vev{\eta_{\perp}^{\mu_1\mu_2}\eta_{\perp}^{\mu_3\mu_4} \eta_{\perp}^{\mu_5\mu_6}\eta_{\perp}^{\mu_7\mu_8}}
    &= 
    a \; \eta_{\perp}^{\mu_1\mu_2}\eta_{\perp}^{\mu_3\mu_4} \eta_{\perp}^{\mu_5\mu_6}\eta_{\perp}^{\mu_7\mu_8}\nonumber\\
    &+ b \,\left( \eta_{\perp}^{\mu_1\mu_3}\eta_{\perp}^{\mu_2\mu_4} \eta_{\perp}^{\mu_5\mu_6}\eta_{\perp}^{\mu_7\mu_8} + ...\right)\nonumber\\
    &+ c \,\left( \eta_{\perp}^{\mu_1\mu_3}\eta_{\perp}^{\mu_2\mu_4} \eta_{\perp}^{\mu_5\mu_7}\eta_{\perp}^{\mu_6\mu_8} + ...\right)\nonumber\\
    &+ d \,\left( \eta_{\perp}^{\mu_1\mu_3}\eta_{\perp}^{\mu_2\mu_5} \eta_{\perp}^{\mu_4\mu_6}\eta_{\perp}^{\mu_7\mu_8} + ...\right)\nonumber\\
    &+ e \,\left( \eta_{\perp}^{\mu_1\mu_3}\eta_{\perp}^{\mu_2\mu_7} \eta_{\perp}^{\mu_4\mu_6}\eta_{\perp}^{\mu_5\mu_8} + ...\right).
    \label{eq:rank8appendix}
\end{align}
Using five contractions with each type of transverse metric tensor product leads to
\begin{align}
    a &= \frac{1}{A} (D_\perp-2)\left(D_\perp +3\right) \left(D_\perp^{2}+6 D_\perp +1\right) \nonumber\\
    b &= \frac{1}{A} \left(-D_\perp^{3}-6 D_\perp^{2}-3 D_\perp +6\right),\nonumber\\
    c &= \frac{1}{A}\left(D_\perp^{2}+5 D_\perp +18\right),\nonumber\\
    d &= \frac{2}{A} D_\perp (D_\perp+4),\nonumber\\
    e &= \frac{1}{A}\left(-5 D_\perp -6\right),
\end{align}
with $A= D_\perp (D_\perp-1)(D_\perp+1) (D_\perp-2)(D_\perp+2)(D_\perp-3)(D_\perp+4)(D_\perp+6)$. 
We notice that one can establish a 1-to-1 correspondence between the unique integer partitions of $R/2=4$ and the possible classes of exchanges between the 4 pairs of indices at rank 8, as shown in Table \ref{tab:my_label}. Following this reasoning, 
we find 7 coefficients at rank 10 (partitions: 5, 4+1, 3+1+1, 3+2, 2+1+1+1, 2+2+1, 1+1+1+1+1), 
then 11 coefficients for rank 12, 15 coefficients for rank 14 and 22 coefficients for rank 16. These coefficients up to rank $14$ are available in the ancillary file, where the coefficients are calculated using \textsc{Form}.

\begin{table}[H]
    \centering
    \begin{tabular}{|c|c|}
    \hline
    Class Representative & Partition of $R/2$ \\ \hline
             $[(12)(34)(56)(78)]$ & 1+1+1+1
             \\ \hline
             $[(1\underparenthesis{ 2 )( 3}4)(56)(78)]$  & 2+1+1
             \\ \hline
             $[(1\underparenthesis{ 2 )( 3}4)(5\underparenthesis{ 6 )( 7}8)]
             $ & 2+2
             \\ \hline   
             $[(1\underparenthesis{ 2 )( 3}\underparenthesis{ 4 )( 5}6)(78)]$ & 3+1
             \\ \hline
     $[(1\underparenthesis{ 2 )( 3}\underparenthesis{ 4 )( 5}\underparenthesis{ 6 )( 7}8)]$ & 4
             \\ \hline
    \end{tabular}
    \caption{Correspondence of class representatives and partitions of $R/2$ for $R=8$.}
    \label{tab:my_label}
\end{table}

\section{Proof of transverse metric contraction}
\label{sec:proofmetric}
In this Appendix we prove Eq.\eqref{eq:metric_contraction_short} from which Eq. \eqref{eq:metric_contraction_long} immediately follows. To prove this relation we start by looking at the left hand side of Eq. \eqref{eq:metric_contraction_short}, which is a tensor with free indices $\rho_{1}\ldots\rho_{2n}$ and composed of building blocks $\eta_{\perp,\rho_i\rho_j}$. We have introduced basis and dual basis decomposition for such tensors above and can rewrite the expression using Eq.~\eqref{eq:TRb} as
\begin{align}
    &\eta_{\perp}^{\mu_1\mu_2}\vev{\eta_{\perp,\mu_1\mu_2}\eta_{\perp,\rho_1\rho_2}\ldots\eta_{\perp,\rho_{2n-1}\rho_{2n}}}\nonumber\\
    &=\sum_{a} \eta_{\perp}^{\mu_1\mu_2}\vev{\eta_{\perp,\mu_1\mu_2}\eta_{\perp,\alpha_1\alpha_2}\ldots\eta_{\perp,\alpha_{2n-1}\alpha_{2n}}}  T_{a}^{\alpha_1\dots \alpha_{2n}}\vev{T_a}_{\rho_1\dots\rho_{2n}} \nonumber \\
    & = \;\eta_{\perp}^{\mu_1\mu_2}\vev{\eta_{\perp,\mu_1\mu_2}\eta_{\perp,\alpha_1\alpha_2}\ldots\eta_{\perp,\alpha_{2n-1}\alpha_{2n}}} \nonumber \\
    & \times \bigg[ \eta_{\perp}^{\alpha_1\alpha_2}\ldots \eta_{\perp}^{\alpha_{2n-1}\alpha_{2n}}\vev{\eta_{\perp,\rho_1\rho_2}\ldots\eta_{\perp,\rho_{2n-1}\rho_{2n}}}\nonumber\\  
    & \hspace{2cm}  +\sum \nolimits_{\substack{\sigma\in \leftindex_2 S_{2n}\\\sigma \neq \textrm{id} }}\eta_{\perp}^{\sigma(\alpha_1)\sigma(\alpha_2)}\ldots \eta_{\perp}^{\sigma(\alpha_{2n-1})\sigma(\alpha_{2n})}\vev{\eta_{\perp,\sigma(\rho_1)\sigma(\rho_2)}\ldots\eta_{\perp,\sigma(\rho_{2n-1})\sigma(\rho_{2n})}} \bigg] \nonumber\\
    &=\, \vev{\eta_{\perp,\rho_1\rho_2}\ldots\eta_{\perp,\rho_{2n-1}\rho_{2n}}} ,
\end{align}
where in the first step we projected onto a basis for rank $2n$ tensors $T_a$. Since the tensor only depends on the transverse metric we can rewrite the expression as done in the second step. We have defined a permutation group $\leftindex_2 S_{2n}$ under which the metric is not invariant. This leads to all independent tensors appearing in the basis, except for the identity which is written separately. In the last step we used the orthogonality conditions for the dual transverse metrics, such that the sum vanishes and only the first term survives. All contractions with a transverse metric not contained in the dual vanish upon contracting with the dual transverse metric product, for instance
$\eta_{\perp}^{\mu_1\rho_1}\vev{\eta_{\perp,\mu_1\mu_2}\eta_{\perp,\rho_1\rho_2}\ldots\eta_{\perp,\rho_{2n-1}\rho_{2n}}}=0.$

\section{One-loop and two-loop $N_f$ coefficients}
\label{sec:Ap_coeff}
In this Appendix, we present for completeness the components of the one-loop amplitude and the $N_f$ contribution to the two-loop amplitude which we computed in Section~\ref{sec:applications}. 
Our results agree with an independent numerical computation by Dario Kermanschah based on the method of Ref.~\cite{Kermanschah:2021wbk}. The full two-loop amplitude for $q \bar q \to \gamma \gamma$ production has been computed analytically in Refs~\cite{Anastasiou:2002zn,Bern:2003ck}.  The corresponding three-loop amplitude has been computed in Ref.~\cite{Bargiela:2021wuy}. 

The one-loop amplitude coefficients for Eq.~\eqref{eq:1loop_result} are
\begin{align}
    h_1^{(1)}(s,t) &=L_y^{2}+\pi^{2}+4 L_s +3 L_y -12,\\
    h_2^{(1)}(s,t) &=  -4 L_s -\frac{\left(x -1\right) L_y^{2}}{x^{2}}-\frac{2 \left(x -1\right) L_y}{x}+12-\frac{\left(x -1\right) \pi^{2}}{x^{2}},\\
    h_3^{(1)}(s,t) &= \frac{4 \left(x -1\right) \left(x +1\right) L_y^{2}}{ \,x^{3}}-\frac{4 \left(x +2\right) L_y}{x^{2} }+\frac{4 \left(x -1\right) \left(x +1\right) \pi^{2}}{ \,x^{3}}-\frac{4}{x},\\
    h_4^{(1)}(s,t) &= 4 L_s +\frac{\left(2 x +1\right) \left(x -1\right) L_y^{2}}{x^{2}}+\frac{2 \left(x -1\right) L_y}{x}-14+\frac{\left(2 x +1\right) \left(x -1\right) \pi^{2}}{x^{2}},
\end{align}
with 
\begin{align}
    x = -\frac{t}{s},\quad L_y = \ln(1-x)+i\pi, \quad L_s = \ln\left(\frac{s}{M^2}\right)-i\pi.
\end{align}

The coefficients of the finite $N_f$ two-loop amplitude displayed in Eq.~\eqref{eq:Nf_result} are
\begin{align}
    h_1^{(2,N_f)}(s,t) &= -\frac{63}{2}-2 L_s^{2}-L_s \,L_y^{2}-3 L_s L_y -L_s \,\pi^{2}+15 L_s-\frac{2 L_y^{3}}{3}  \nonumber\\
    &\hspace{1em}+\frac{\left(5 x -4\right) L_y^{2}}{2 x}-\frac{2 L_y \,\pi^{2}}{3}+\frac{21 L_y}{2}+\frac{\left(9 x -4\right) \pi^{2}}{2 x},\\
    h_2^{(2,N_f)}(s,t) &=  2 L_s^{2}+\frac{\left(x -1\right) L_s \,L_y^{2}}{x^{2}}+\frac{2 \left(x -1\right) L_s L_y}{x}+\frac{\left(x -1\right) L_s \,\pi^{2}}{x^{2}}
    -14 L_s  \nonumber\\
    &\hspace{1em}+\frac{2 \left(x -1\right) L_y^{3}}{3 x^{2}}-\frac{7 \left(x -1\right) L_y^{2}}{2 x^{2}}+\frac{2 \left(x -1\right) L_y \,\pi^{2}}{3 x^{2}} 
    -\frac{7 \left(x -1\right) L_y}{x}\nonumber\\
    &\hspace{1em}+29-\frac{\left(8 x +21\right) \left(x -1\right) \pi^{2}}{6 x^{2}},\\
    h_3^{(2,N_f)}(s,t) &=  -\frac{4 \left(x -1\right) \left(x +1\right) L_s \,L_y^{2}}{ \,x^{3}}+\frac{4 \left(x +2\right) L_s L_y}{x^{2} }   \nonumber\\
    &\hspace{1em}
    -\frac{4 \left(x -1\right) \left(x +1\right) L_s \,\pi^{2}}{\,x^{3}}+\frac{4 L_s}{x }-\frac{8 \left(x -1\right) \left(x +1\right) L_y^{3}}{3  \,x^{3}} 
    \nonumber\\
    &\hspace{1em}
    +\frac{2 \left(9 x^{2}+4 x -7\right) L_y^{2}}{ \,x^{3}}-\frac{8 \left(x -1\right) \left(x +1\right) L_y \,\pi^{2}}{3  \,x^{3}}-\frac{2 \left(3 x +14\right) L_y}{x^{2}}  
    \nonumber\\
    &\hspace{1em}
    +\frac{2 \left(23 x^{2}+4 x -21\right) \pi^{2}}{3  \,x^{3}}-\frac{14}{x },\\
    h_4^{(2,N_f)}(s,t) &= -2 L_s^{2}-\frac{\left(2 x +1\right) \left(x -1\right) L_s \,L_y^{2}}{x^{2}}   -\frac{2 \left(x -1\right) L_s L_y}{x} 
    \nonumber\\
    &\hspace{1em} 
    -\frac{\left(2 x +1\right) \left(x -1\right) L_s \,\pi^{2}}{x^{2}}+16 L_s -\frac{2 \left(2 x +1\right) \left(x -1\right) L_y^{3}}{3 x^{2}}
     \nonumber\\
    &\hspace{1em}
    +\frac{7 \left(2 x +1\right) \left(x -1\right) L_y^{2}}{2 x^{2}} -\frac{2 \left(2 x +1\right) \left(x -1\right) L_y \,\pi^{2}}{3 x^{2}}+\frac{\left(9 x -7\right) L_y}{x}
     \nonumber\\
    &\hspace{1em}
    -36 +\frac{\left(50 x +21\right) \left(x -1\right) \pi^{2}}{6 x^{2}}.
\end{align}


\begin{thebibliography}{10}

\bibitem{TKACHOV198165}
F.~Tkachov, {\it A theorem on analytical calculability of 4-loop
  renormalization group functions},  {\em Physics Letters B} {\bf 100} (1981),
  no.~1 65--68.

\bibitem{CHETYRKIN1981159}
K.~Chetyrkin and F.~Tkachov, {\it Integration by parts: The algorithm to
  calculate $\beta$-functions in 4 loops},  {\em Nuclear Physics B} {\bf 192}
  (1981), no.~1 159--204.

\bibitem{Laporta:2001dd}
S.~Laporta, {\it {High precision calculation of multiloop Feynman integrals by
  difference equations}},  {\em Int. J. Mod. Phys. A} {\bf 15} (2000)
  5087--5159, [\href{http://arxiv.org/abs/hep-ph/0102033}{{\tt
  hep-ph/0102033}}].

\bibitem{Passarino:1978jh}
G.~Passarino and M.~J.~G. Veltman, {\it {One Loop Corrections for e+ e-
  Annihilation Into mu+ mu- in the Weinberg Model}},  {\em Nucl. Phys. B} {\bf
  160} (1979) 151--207.

\bibitem{Ezawa:1990dh}
Y.~Ezawa {\em et~al.}, {\it {Brown-Feynman reduction of one loop Feynman
  diagrams to scalar integrals with orthonormal basis tensors}},  {\em Comput.
  Phys. Commun.} {\bf 69} (1992) 15--45.

\bibitem{Denner:2005nn}
A.~Denner and S.~Dittmaier, {\it {Reduction schemes for one-loop tensor
  integrals}},  {\em Nucl. Phys. B} {\bf 734} (2006) 62--115,
  [\href{http://arxiv.org/abs/hep-ph/0509141}{{\tt hep-ph/0509141}}].

\bibitem{Binoth:2008uq}
T.~Binoth, J.~P. Guillet, G.~Heinrich, E.~Pilon, and T.~Reiter, {\it {Golem95:
  A Numerical program to calculate one-loop tensor integrals with up to six
  external legs}},  {\em Comput. Phys. Commun.} {\bf 180} (2009) 2317--2330,
  [\href{http://arxiv.org/abs/0810.0992}{{\tt 0810.0992}}].

\bibitem{vanHameren:2009vq}
A.~van Hameren, {\it {Multi-gluon one-loop amplitudes using tensor integrals}},
   {\em JHEP} {\bf 07} (2009) 088, [\href{http://arxiv.org/abs/0905.1005}{{\tt
  0905.1005}}].

\bibitem{vanHameren:2009dr}
A.~van Hameren, C.~G. Papadopoulos, and R.~Pittau, {\it {Automated one-loop
  calculations: A Proof of concept}},  {\em JHEP} {\bf 09} (2009) 106,
  [\href{http://arxiv.org/abs/0903.4665}{{\tt 0903.4665}}].

\bibitem{Cascioli:2011va}
F.~Cascioli, P.~Maierhofer, and S.~Pozzorini, {\it {Scattering Amplitudes with
  Open Loops}},  {\em Phys. Rev. Lett.} {\bf 108} (2012) 111601,
  [\href{http://arxiv.org/abs/1111.5206}{{\tt 1111.5206}}].

\bibitem{delAguila:2004nf}
F.~del Aguila and R.~Pittau, {\it {Recursive numerical calculus of one-loop
  tensor integrals}},  {\em JHEP} {\bf 07} (2004) 017,
  [\href{http://arxiv.org/abs/hep-ph/0404120}{{\tt hep-ph/0404120}}].

\bibitem{Britto:2004nc}
R.~Britto, F.~Cachazo, and B.~Feng, {\it {Generalized unitarity and one-loop
  amplitudes in N=4 super-Yang-Mills}},  {\em Nucl. Phys. B} {\bf 725} (2005)
  275--305, [\href{http://arxiv.org/abs/hep-th/0412103}{{\tt hep-th/0412103}}].

\bibitem{Britto:2006sj}
R.~Britto, B.~Feng, and P.~Mastrolia, {\it {The Cut-constructible part of QCD
  amplitudes}},  {\em Phys. Rev. D} {\bf 73} (2006) 105004,
  [\href{http://arxiv.org/abs/hep-ph/0602178}{{\tt hep-ph/0602178}}].

\bibitem{Ossola:2006us}
G.~Ossola, C.~G. Papadopoulos, and R.~Pittau, {\it {Reducing full one-loop
  amplitudes to scalar integrals at the integrand level}},  {\em Nucl. Phys. B}
  {\bf 763} (2007) 147--169, [\href{http://arxiv.org/abs/hep-ph/0609007}{{\tt
  hep-ph/0609007}}].

\bibitem{Forde:2007mi}
D.~Forde, {\it {Direct extraction of one-loop integral coefficients}},  {\em
  Phys. Rev. D} {\bf 75} (2007) 125019,
  [\href{http://arxiv.org/abs/0704.1835}{{\tt 0704.1835}}].

\bibitem{Ellis:2007br}
R.~K. Ellis, W.~T. Giele, and Z.~Kunszt, {\it {A Numerical Unitarity Formalism
  for Evaluating One-Loop Amplitudes}},  {\em JHEP} {\bf 03} (2008) 003,
  [\href{http://arxiv.org/abs/0708.2398}{{\tt 0708.2398}}].

\bibitem{Giele:2008bc}
W.~T. Giele and G.~Zanderighi, {\it {On the Numerical Evaluation of One-Loop
  Amplitudes: The Gluonic Case}},  {\em JHEP} {\bf 06} (2008) 038,
  [\href{http://arxiv.org/abs/0805.2152}{{\tt 0805.2152}}].

\bibitem{Lazopoulos:2008ex}
A.~Lazopoulos, {\it {Multi-gluon one-loop amplitudes numerically}},
  \href{http://arxiv.org/abs/0812.2998}{{\tt 0812.2998}}.

\bibitem{Berger:2008sj}
C.~F. Berger, Z.~Bern, L.~J. Dixon, F.~Febres~Cordero, D.~Forde, H.~Ita, D.~A.
  Kosower, and D.~Maitre, {\it {An Automated Implementation of On-Shell Methods
  for One-Loop Amplitudes}},  {\em Phys. Rev. D} {\bf 78} (2008) 036003,
  [\href{http://arxiv.org/abs/0803.4180}{{\tt 0803.4180}}].

\bibitem{Ellis:2008ir}
R.~K. Ellis, W.~T. Giele, Z.~Kunszt, and K.~Melnikov, {\it {Masses, fermions
  and generalized $D$-dimensional unitarity}},  {\em Nucl. Phys. B} {\bf 822}
  (2009) 270--282, [\href{http://arxiv.org/abs/0806.3467}{{\tt 0806.3467}}].

\bibitem{Berger:2009zg}
C.~F. Berger, Z.~Bern, L.~J. Dixon, F.~Febres~Cordero, D.~Forde, T.~Gleisberg,
  H.~Ita, D.~A. Kosower, and D.~Maitre, {\it {Precise Predictions for $W$ + 3
  Jet Production at Hadron Colliders}},  {\em Phys. Rev. Lett.} {\bf 102}
  (2009) 222001, [\href{http://arxiv.org/abs/0902.2760}{{\tt 0902.2760}}].

\bibitem{Berger:2009zb}
C.~F. Berger and D.~Forde, {\it {Multi-Parton Scattering Amplitudes via
  On-Shell Methods}},  {\em Ann. Rev. Nucl. Part. Sci.} {\bf 60} (2010)
  181--205, [\href{http://arxiv.org/abs/0912.3534}{{\tt 0912.3534}}].

\bibitem{Britto:2010xq}
R.~Britto, {\it {Loop Amplitudes in Gauge Theories: Modern Analytic
  Approaches}},  {\em J. Phys. A} {\bf 44} (2011) 454006,
  [\href{http://arxiv.org/abs/1012.4493}{{\tt 1012.4493}}].

\bibitem{Ellis:2011cr}
R.~K. Ellis, Z.~Kunszt, K.~Melnikov, and G.~Zanderighi, {\it {One-loop
  calculations in quantum field theory: from Feynman diagrams to unitarity
  cuts}},  {\em Phys. Rept.} {\bf 518} (2012) 141--250,
  [\href{http://arxiv.org/abs/1105.4319}{{\tt 1105.4319}}].

\bibitem{Actis:2012qn}
S.~Actis, A.~Denner, L.~Hofer, A.~Scharf, and S.~Uccirati, {\it {Recursive
  generation of one-loop amplitudes in the Standard Model}},  {\em JHEP} {\bf
  04} (2013) 037, [\href{http://arxiv.org/abs/1211.6316}{{\tt 1211.6316}}].

\bibitem{Ita:2015tya}
H.~Ita, {\it {Two-loop Integrand Decomposition into Master Integrals and
  Surface Terms}},  {\em Phys. Rev. D} {\bf 94} (2016), no.~11 116015,
  [\href{http://arxiv.org/abs/1510.05626}{{\tt 1510.05626}}].

\bibitem{Abreu:2017hqn}
S.~Abreu, F.~Febres~Cordero, H.~Ita, B.~Page, and M.~Zeng, {\it {Planar
  Two-Loop Five-Gluon Amplitudes from Numerical Unitarity}},  {\em Phys. Rev.
  D} {\bf 97} (2018), no.~11 116014,
  [\href{http://arxiv.org/abs/1712.03946}{{\tt 1712.03946}}].

\bibitem{Badger:2017jhb}
S.~Badger, C.~Br\o{}nnum-Hansen, H.~B. Hartanto, and T.~Peraro, {\it {First
  look at two-loop five-gluon scattering in QCD}},  {\em Phys. Rev. Lett.} {\bf
  120} (2018), no.~9 092001, [\href{http://arxiv.org/abs/1712.02229}{{\tt
  1712.02229}}].

\bibitem{Abreu:2020lyk}
S.~Abreu, F.~Febres~Cordero, H.~Ita, M.~Jaquier, B.~Page, M.~S. Ruf, and
  V.~Sotnikov, {\it {Two-Loop Four-Graviton Scattering Amplitudes}},  {\em
  Phys. Rev. Lett.} {\bf 124} (2020), no.~21 211601,
  [\href{http://arxiv.org/abs/2002.12374}{{\tt 2002.12374}}].

\bibitem{Abreu:2020xvt}
S.~Abreu, J.~Dormans, F.~Febres~Cordero, H.~Ita, M.~Kraus, B.~Page, E.~Pascual,
  M.~S. Ruf, and V.~Sotnikov, {\it {Caravel: A C++ framework for the
  computation of multi-loop amplitudes with numerical unitarity}},  {\em
  Comput. Phys. Commun.} {\bf 267} (2021) 108069,
  [\href{http://arxiv.org/abs/2009.11957}{{\tt 2009.11957}}].

\bibitem{Abreu:2021asb}
S.~Abreu, F.~Febres~Cordero, H.~Ita, M.~Klinkert, B.~Page, and V.~Sotnikov,
  {\it {Leading-color two-loop amplitudes for four partons and a W boson in
  QCD}},  {\em JHEP} {\bf 04} (2022) 042,
  [\href{http://arxiv.org/abs/2110.07541}{{\tt 2110.07541}}].

\bibitem{Badger:2022ncb}
S.~Badger, H.~B. Hartanto, J.~Kry\'s, and S.~Zoia, {\it {Two-loop leading
  colour helicity amplitudes for W$^{±}$\ensuremath{\gamma} + j production at
  the LHC}},  {\em JHEP} {\bf 05} (2022) 035,
  [\href{http://arxiv.org/abs/2201.04075}{{\tt 2201.04075}}].

\bibitem{Abreu:2023bdp}
S.~Abreu, G.~De~Laurentis, H.~Ita, M.~Klinkert, B.~Page, and V.~Sotnikov, {\it
  {Two-Loop QCD Corrections for Three-Photon Production at Hadron Colliders}},
  \href{http://arxiv.org/abs/2305.17056}{{\tt 2305.17056}}.

\bibitem{Badger:2023mgf}
S.~Badger, M.~Czakon, H.~B. Hartanto, R.~Moodie, T.~Peraro, R.~Poncelet, and
  S.~Zoia, {\it {Isolated photon production in association with a jet pair
  through next-to-next-to-leading order in QCD}},
  \href{http://arxiv.org/abs/2304.06682}{{\tt 2304.06682}}.

\bibitem{Tarasov:1996br}
O.~V. Tarasov, {\it {Connection between Feynman integrals having different
  values of the space-time dimension}},  {\em Phys. Rev. D} {\bf 54} (1996)
  6479--6490, [\href{http://arxiv.org/abs/hep-th/9606018}{{\tt
  hep-th/9606018}}].

\bibitem{Anastasiou:1999bn}
C.~Anastasiou, E.~W.~N. Glover, and C.~Oleari, {\it {The two-loop scalar and
  tensor pentabox graph with light-like legs}},  {\em Nucl. Phys. B} {\bf 575}
  (2000) 416--436, [\href{http://arxiv.org/abs/hep-ph/9912251}{{\tt
  hep-ph/9912251}}]. [Erratum: Nucl.Phys.B 585, 763--770 (2000)].

\bibitem{Chen:2019wyb}
L.~Chen, {\it {A prescription for projectors to compute helicity amplitudes in
  D dimensions}},  {\em Eur. Phys. J. C} {\bf 81} (2021), no.~5 417,
  [\href{http://arxiv.org/abs/1904.00705}{{\tt 1904.00705}}].

\bibitem{Peraro:2019cjj}
T.~Peraro and L.~Tancredi, {\it {Physical projectors for multi-leg helicity
  amplitudes}},  {\em JHEP} {\bf 07} (2019) 114,
  [\href{http://arxiv.org/abs/1906.03298}{{\tt 1906.03298}}].

\bibitem{Peraro:2020sfm}
T.~Peraro and L.~Tancredi, {\it {Tensor decomposition for bosonic and fermionic
  scattering amplitudes}},  {\em Phys. Rev. D} {\bf 103} (2021), no.~5 054042,
  [\href{http://arxiv.org/abs/2012.00820}{{\tt 2012.00820}}].

\bibitem{Garland:2002ak}
L.~W. Garland, T.~Gehrmann, E.~W.~N. Glover, A.~Koukoutsakis, and E.~Remiddi,
  {\it {Two loop QCD helicity amplitudes for e+ e- ---\ensuremath{>} three
  jets}},  {\em Nucl. Phys. B} {\bf 642} (2002) 227--262,
  [\href{http://arxiv.org/abs/hep-ph/0206067}{{\tt hep-ph/0206067}}].

\bibitem{Bern:2002tk}
Z.~Bern, A.~De~Freitas, and L.~J. Dixon, {\it {Two loop helicity amplitudes for
  gluon-gluon scattering in QCD and supersymmetric Yang-Mills theory}},  {\em
  JHEP} {\bf 03} (2002) 018, [\href{http://arxiv.org/abs/hep-ph/0201161}{{\tt
  hep-ph/0201161}}].

\bibitem{Caola:2022dfa}
F.~Caola, A.~Chakraborty, G.~Gambuti, A.~von Manteuffel, and L.~Tancredi, {\it
  {Three-loop helicity amplitudes for quark-gluon scattering in QCD}},  {\em
  JHEP} {\bf 12} (2022) 082, [\href{http://arxiv.org/abs/2207.03503}{{\tt
  2207.03503}}].

\bibitem{Badger:2023xtl}
S.~Badger, J.~Kry\'s, R.~Moodie, and S.~Zoia, {\it {Lepton-pair scattering with
  an off-shell and an on-shell photon at two loops in massless QED}},
  \href{http://arxiv.org/abs/2307.03098}{{\tt 2307.03098}}.

\bibitem{Gehrmann:2023jyv}
T.~Gehrmann, P.~Jakub\v{c}\'\i{}k, C.~C. Mella, N.~Syrrakos, and L.~Tancredi,
  {\it {Planar three-loop QCD helicity amplitudes for $V$+jet production at
  hadron colliders}},  \href{http://arxiv.org/abs/2307.15405}{{\tt
  2307.15405}}.

\bibitem{Gehrmann:2023zpz}
T.~Gehrmann, P.~Jakub\v{c}\'\i{}k, C.~C. Mella, N.~Syrrakos, and L.~Tancredi,
  {\it {Two-loop helicity amplitudes for $V+$jet production including axial
  vector couplings to higher orders in $\epsilon$}},
  \href{http://arxiv.org/abs/2306.10170}{{\tt 2306.10170}}.

\bibitem{Anastasiou:2022udy}
C.~Anastasiou, D.~P.~L. Bragan\c{c}a, L.~Senatore, and H.~Zheng, {\it
  {Efficiently evaluating loop integrals in the EFTofLSS using QFT integrals
  with massive propagators}},  \href{http://arxiv.org/abs/2212.07421}{{\tt
  2212.07421}}.

\bibitem{Ruijl:2018poj}
B.~Ruijl, F.~Herzog, T.~Ueda, J.~A.~M. Vermaseren, and A.~Vogt, {\it {The
  Rstar-operation and five-loop calculations}},  {\em PoS} {\bf RADCOR2017}
  (2018) 011, [\href{http://arxiv.org/abs/1801.06084}{{\tt 1801.06084}}].

\bibitem{vanNeerven:1983vr}
W.~L. van Neerven and J.~A.~M. Vermaseren, {\it {LARGE LOOP INTEGRALS}},  {\em
  Phys. Lett. B} {\bf 137} (1984) 241--244.

\bibitem{vermaseren2000new}
J.~A.~M. Vermaseren, {\it New features of form},  2000.

\bibitem{Tentyukov_2010}
M.~Tentyukov and J.~Vermaseren, {\it The multithreaded version of {FORM}},
  {\em Computer Physics Communications} {\bf 181} (aug, 2010) 1419--1427.

\bibitem{Kuipers_2013}
J.~Kuipers, T.~Ueda, J.~Vermaseren, and J.~Vollinga, {\it {FORM} version 4.0},
  {\em Computer Physics Communications} {\bf 184} (may, 2013) 1453--1467.

\bibitem{Smith:2004ck}
J.~Smith and W.~L. van Neerven, {\it {The Difference between
    n-dimensional regularization and n-dimensional reduction in QCD}},  {\em Eur. Phys. J. C}
  {\bf 40} (2005) 199--203, [\href{http://arxiv.org/abs/hep-ph/0411357}{{\tt
  hep-ph/0411357}}].

\bibitem{Anastasiou:2022eym}
C.~Anastasiou and G.~Sterman, {\it {Locally finite two-loop QCD amplitudes from
  IR universality for electroweak production}},  {\em JHEP} {\bf 05} (2023)
  242, [\href{http://arxiv.org/abs/2212.12162}{{\tt 2212.12162}}].

\bibitem{Anastasiou:2004vj}
C.~Anastasiou and A.~Lazopoulos, {\it {Automatic integral reduction for higher
  order perturbative calculations}},  {\em JHEP} {\bf 07} (2004) 046,
  [\href{http://arxiv.org/abs/hep-ph/0404258}{{\tt hep-ph/0404258}}].

\bibitem{Anastasiou:2002zn}
C.~Anastasiou, E.~W.~N. Glover, and M.~E. Tejeda-Yeomans, {\it {Two loop QED
  and QCD corrections to massless fermion boson scattering}},  {\em Nucl. Phys.
  B} {\bf 629} (2002) 255--289,
  [\href{http://arxiv.org/abs/hep-ph/0201274}{{\tt hep-ph/0201274}}].

\bibitem{Anastasiou:2020sdt}
C.~Anastasiou, R.~Haindl, G.~Sterman, Z.~Yang, and M.~Zeng, {\it {Locally
  finite two-loop amplitudes for off-shell multi-photon production in
  electron-positron annihilation}},  {\em JHEP} {\bf 04} (2021) 222,
  [\href{http://arxiv.org/abs/2008.12293}{{\tt 2008.12293}}].

\bibitem{Tausk:1999vh}
J.~B. Tausk, {\it {Nonplanar massless two loop Feynman diagrams with four
  on-shell legs}},  {\em Phys. Lett. B} {\bf 469} (1999) 225--234,
  [\href{http://arxiv.org/abs/hep-ph/9909506}{{\tt hep-ph/9909506}}].

\bibitem{Smirnov:1999gc}
V.~A. Smirnov, {\it {Analytical result for dimensionally regularized massless
  on shell double box}},  {\em Phys. Lett. B} {\bf 460} (1999) 397--404,
  [\href{http://arxiv.org/abs/hep-ph/9905323}{{\tt hep-ph/9905323}}].

\bibitem{Anastasiou:2000kp}
C.~Anastasiou, J.~B. Tausk, and M.~E. Tejeda-Yeomans, {\it {The On-shell
  massless planar double box diagram with an irreducible numerator}},  {\em
  Nucl. Phys. B Proc. Suppl.} {\bf 89} (2000) 262--267,
  [\href{http://arxiv.org/abs/hep-ph/0005328}{{\tt hep-ph/0005328}}].

\bibitem{Anastasiou:2018rib}
C.~Anastasiou and G.~Sterman, {\it {Removing infrared divergences from two-loop
  integrals}},  {\em JHEP} {\bf 07} (2019) 056,
  [\href{http://arxiv.org/abs/1812.03753}{{\tt 1812.03753}}].

\bibitem{Kermanschah:2021wbk}
D.~Kermanschah, {\it {Numerical integration of loop integrals through local
  cancellation of threshold singularities}},  {\em JHEP} {\bf 01} (2022) 151,
  [\href{http://arxiv.org/abs/2110.06869}{{\tt 2110.06869}}].

\bibitem{Bern:2003ck}
Z.~Bern, A.~De~Freitas, and L.~J. Dixon, {\it {Two loop helicity amplitudes for
  quark gluon scattering in QCD and gluino gluon scattering in supersymmetric
  Yang-Mills theory}},  {\em JHEP} {\bf 06} (2003) 028,
  [\href{http://arxiv.org/abs/hep-ph/0304168}{{\tt hep-ph/0304168}}]. [Erratum:
  JHEP 04, 112 (2014)].

\bibitem{Bargiela:2021wuy}
P.~Bargiela, F.~Caola, A.~von Manteuffel, and L.~Tancredi, {\it {Three-loop
  helicity amplitudes for diphoton production in gluon fusion}},  {\em JHEP}
  {\bf 02} (2022) 153, [\href{http://arxiv.org/abs/2111.13595}{{\tt
  2111.13595}}].

\end{thebibliography}

\providecommand{\href}[2]{#2}\begingroup\raggedright\endgroup

\end{document}